\documentclass[12pt,final]{myelsart}

  \newcommand{\preprint}{ECT*--06--09\\SACLAY--T06/067}

  \usepackage{latexsym,bm,amsmath,amssymb,amsfonts}
  \usepackage{epsfig,graphics,graphicx}
  \usepackage{cite}
  \usepackage{slashed}

  \renewcommand{\theequation}{\thesection.\arabic{equation}}
  \long\def\comment#1{ }

  \newcommand{\dif}{{\rm d}}
  \newcommand{\dY}{\dif Y}
  \newcommand{\abar}{\bar{\alpha}_s}
  
  \newcommand{\del}{\partial}
  \newcommand{\lan}{\langle}
  \newcommand{\ran}{\rangle}
  \newcommand{\mcal}{\mathcal}
  \newcommand{\rme}{{\rm e}}
  \newcommand{\rmi}{{\rm i}}
  \newcommand{\rmL}{{\rm L}}
  \newcommand{\rmR}{{\rm R}}

  \newcommand{\wt}{\widetilde}

  \newcommand{\nn}{\nonumber\\}
   \newcommand{\wtP}{\widetilde{P}}
  \newcommand{\order}[1]{\mcal{O}{(#1)}}

  \newcommand{\beq}{\begin{eqnarray}}
  \newcommand{\eeq}{\end{eqnarray}}
  \newcommand{\avg}[1]{\left\langle #1 \right\rangle}
 \def\simge{\mathrel{%
   \rlap{\raise 0.511ex \hbox{$>$}}{\lower 0.511ex \hbox{$\sim$}}}}
\def\simle{\mathrel{
   \rlap{\raise 0.511ex \hbox{$<$}}{\lower 0.511ex \hbox{$\sim$}}}}

\begin{document}

\begin{flushright}
{\small \preprint}
\end{flushright}
\vspace{1.5cm}

\begin{frontmatter}\parbox[]{16.0cm}{\begin{center}
\title{\rm \LARGE A zero--dimensional model for high--energy\\
scattering in QCD}

\author{J.-P.~Blaizot $^{\rm a,1,}$\thanksref{th2}},
\author{E.~Iancu $^{\rm b,1,}$\thanksref{th2}},
\author{D.N.~Triantafyllopoulos $^{\rm a,1}$}

\address{$^{\rm a}$ ECT$^*$, Villa Tambosi, Strada delle Tabarelle
286, I-38050 Villazzano (TN), Italy}

\address{$^{\rm b}$ Service de Physique Theorique, CEA Saclay,
CEA/DSM/SPhT, F-91191 Gif-sur-Yvette, France}
\thanks{{\it E-mail addresses:}
blaizot@ect.it (J.-P.~Blaizot), iancu@dsm-mail.cea.fr (E.~Iancu),
dionysis@ect.it (D.N.~Triantafyllopoulos).}
\thanks[th2]{Membre du Centre National de la Recherche Scientifique
(CNRS), France.}

{\small \today}
\begin{abstract}
We investigate a zero--dimensional toy model originally introduced
by Mueller and Salam \cite{AMSalam95} which mimics high--energy
scattering in QCD in the presence of both gluon saturation and gluon
number fluctuations, and hence of Pomeron loops. Unlike other toy
models of the reaction--diffusion type, the model studied in this
paper is consistent with boost invariance and, related to that, it
exhibits a mechanism for particle saturation close to that of the
JIMWLK equation in QCD, namely the saturation of the emission rate
due to high--density effects. Within this model, we establish the
dominant high--energy behaviour of the $S$--matrix element $\langle
S^n \rangle$ for the scattering between a target obtained by
evolving one particle and a projectile made with exactly $n$
particles. Remarkably, we find that all such matrix elements
approach the black disk limit $S=0$ at high rapidity $Y$, with the
same exponential law: $\langle S^n \rangle\sim \exp(-Y)$ for all
values of $n$. This is so because the $S$--matrix is dominated by
{\em rare} target configurations which involve only few particles.
We also find that the bulk distribution for a saturated system is of
the Poisson type.

\end{abstract}
\end{center}}

\end{frontmatter}

\newpage

\section{Introduction}
\setcounter{equation}{0} \label{SECT_INTRO}

Much of the recent progress in the field of high--energy QCD has
come from the gradual understanding
\cite{AM94,AM95,Salam95,B,K,JKLW,W,CGC,PATH,IM031,IM032,MP03,MS04,
IMM04,LL04,IT04,IT05,MSW05,LL05,KL05,KL3,BIIT05,BREM,Balit05,IST05}
of the analogies between the gluon evolution in QCD at high energy
and a classical stochastic process, similar to the
`reaction--diffusion process' $A \rightleftharpoons AA$
\cite{Gardiner,Saar}. Such analogies lead to the conclusion that the
properties of the QCD scattering amplitudes at very high energy,
including in the vicinity of the unitarity limit, are strongly
influenced by {\em gluon--number fluctuations} in the dilute regime,
and hence cannot be reliably computed from mean field approximations
like the Balitsky--Kovchegov (BK) equation \cite{B,K}. Although at a
first sight surprising --- since the high--energy regime is
characterized by high gluon occupancy, and therefore should be less
affected by fluctuations ---, such a strong sensitivity to
fluctuations was in fact noticed in early studies of unitarization
in the context of the dipole picture \cite{AM94,AM95,Salam95} and,
more recently, it has been rediscovered within the context of the
non--linear QCD evolution in the vicinity of the saturation line
\cite{IM032,MS04}.

It has been realized \cite{IT04} that the relevant fluctuations are
missed by the JIMWLK equation \cite{JKLW,W,CGC}, which describes the
non--linear evolution of the gluon distribution in the
(high--density) target, as well as by its `dual' counterpart, the
Balitsky hierarchy \cite{B} for the scattering amplitudes, in which
the evolution is rather encoded in the wavefunction of the (dilute)
projectile. The JIMWLK equation properly encompasses the mechanism
responsible for gluon saturation at high density, but it misses the
correlations associated with gluon splitting in the dilute regime.
Correspondingly, the Balitsky equations faithfully describe the
splitting processes within the projectile, but completely ignore the
non--linear effects responsible for saturation.

At the same time, new equations have been proposed
\cite{IT04,MSW05,IT05,LL05}, which heuristically combine the dipole
picture in the dilute regime with the JIMWLK evolution at high
density, thus leading to a generalization of the Balitsky hierarchy
which encompasses both saturation and fluctuation effects in the
limit where the number of colors $N_c$ is large, $N_c\gg 1$. These
equations have been interpreted \cite{BIIT05} as an effective theory
for BFKL `pomerons', in which the pomerons are allowed to dissociate
and recombine with each other, like the `molecules' in the
reaction--diffusion problem. (See also Ref. \cite{Braun05} for a
related approach in the context of nucleus--nucleus scattering.) But
the complexity of these `Pomeron loop' equations has so far hindered
any systematic approach towards their solutions, including via
numerical methods. The only properties of these solutions to be
presently known come essentially from the correspondence with
statistical physics \cite{MP03,MS04,IMM04,IT04,GS05,EGBM05}, which
is however limited to asymptotically high energies and very small
values of the coupling constant.

In view of the complexity of the QCD problem, several authors
\cite{IT04,Stasto05,KL5,SX05,KozLev06} have recently started
investigating simpler models in  zero transverse dimensions that
allow for more direct studies (often analytic) of the approach
towards saturation and unitarization with increasing energy. These
models are usually formulated in terms of stochastic distribution of
classical, point--like, particles which evolve in `time' (rapidity)
according to some suitable master equation. The models discussed in
Refs. \cite{IT04,Stasto05,SX05,KozLev06} are all borrowed from
statistical physics and describe a $A \rightleftharpoons AA$
reaction process. The model briefly discussed in Ref. \cite{KL5} has
been originally proposed by Mueller and Salam \cite{AMSalam95} as a
toy model for `dipole' evolution in the presence of saturation, and
seems to be the closest to the actual QCD dynamics, as we shall
explain at length in what follows. This is the model that we shall
focus on in this paper.

At this point, one may legitimately wonder about the usefulness of
{\em any} zero--dimensional model in the context of the high--energy
evolution in QCD. As well known, this evolution is genuinely
non--local in the transverse plane, and this non--locality is
responsible for the main characteristic of a hadron wavefunction at
sufficiently high energy, namely, the coexistence of two different
phases at different values of the gluon transverse momentum
$k_\perp$: \texttt{(i)} a dilute phase at large $k_\perp$, where the
standard, linear (or `leading--twist'), perturbative evolution
applies and \texttt{(ii)} a high--density phase, also known as the
{\em color glass condensate} \cite{MV,CGC,EdiCGC}, at relatively low
$k_\perp$, where the dynamics is fully non--linear and the gluon
occupation factor (almost) saturates. With increasing energy, the
CGC phase extends towards larger values of $k_\perp$, and the main
physical questions refer to the rate of this expansion and to the
properties of the gluon distribution and of the scattering
amplitudes in the vicinity of the {\em saturation line}  --- i.e.,
in the transition region from dense to dilute
\cite{SCALING,MT02,DT02}. In particular, it is precisely in this
region that the effects of the gluon--number fluctuations are most
striking \cite{MS04,IMM04,IT04} : they considerably slow down the
energy increase of the saturation momentum and render the saturation
line diffuse, which in turn has important consequences for the
measured cross--sections \cite{HIMST06,GLUON}.

Clearly, all these interesting features are lost when one restricts
oneself to a zero--dimensional problem. But even in that case, some
non--trivial questions remain, whose detailed understanding could
shed light on the physics of saturation in the presence of
particle--number fluctuations. Chiefly among these, there is the
problem of the approach of the (average) $S$--matrix element
$\langle S\rangle$ towards the black disk limit $\langle S\rangle=0$
with increasing energy. The arguments in Refs.
\cite{AMSalam95,IM032} as well as the analysis to be presented in
this paper demonstrate that this approach is strongly influenced by
the fluctuations (at least in specific frames, since the physically
relevant configurations are frame--dependent): the average
$S$--matrix near the unitarity limit is dominated by {\em atypical}
configurations, which have a relatively low gluon occupancy --- well
below the average occupancy at that energy --- and yield a large
contribution $S\sim 1$ to the average quantity $\langle S\rangle$.

The dominance of rare fluctuations on the average $S$--matrix has an
interesting implication\footnote{To our knowledge, this implication
has not been previously noticed in the literature.} for the matrix
element $\langle S^n\rangle$ of a projectile made with $n$
particles. As well known, the quantity $|\langle S^n\rangle|^2$
measures the probability that the system of $n$ particles emerge
unscattered. Simple arguments of the mean--field type suggest that
$\langle S^n\rangle\sim \langle S\rangle^n$ at high energy. (For
instance, this is the prediction of the Balitsky--JIMWLK equations.)
However, if the average $S$--matrix is dominated by {\em rare}
configurations for which $S\sim 1$ per incoming particle, then for
these relevant configurations we have $S^n \sim S \sim 1$ for any
$n$, which after averaging yields $\langle S^n\rangle  \sim \langle
S\rangle$, in sharp contrast with the mean field expectation. These
arguments will be confirmed by our subsequent analysis of the
Mueller--Salam model; for instance, in the case where the target is
obtained by evolving a single particle, we shall find that $\langle
S^n \rangle\sim \exp(-Y)$ for any $n$.

Let us now explain why, in our opinion, the model introduced by
Mueller and Salam in Ref. \cite{AMSalam95} is indeed closer to the
actual QCD dynamics than the (zero--dimensional) reaction model $A
\rightleftharpoons AA$ (although the latter appears to more widely
studied in the recent literature). The main reason is that the
Mueller--Salam model correctly captures the QCD mechanism for the
saturation of the particle number at high energy, which is {\em not}
particle recombination (as in the reaction--diffusion process), but
rather the {\em saturation of the rate for particle emission due to
high--density effects}. The new particle which is emitted when
increasing the rapidity in one step can be radiated off any of the
particles created in the previous steps and, moreover, it can
undergo multiple scattering off the latter. When the density is
relatively low (the low--energy case), the emission rate is simply
proportional to the total number of sources, leading to an
exponential increase of the number of particles with the rapidity
$Y$. But when the density becomes large enough for multiple
scattering to be important, the emission rate saturates at a
constant value, independent of the number of particles in the
system. Then, the number of particles keeps growing, but only {\em
linearly} in $Y$. This `almost saturation' scenario is indeed
similar to the physical mechanism at work in QCD \cite{AM99,SAT},
where gluon saturation
--- as described by the JIMWLK equation --- proceeds via
the saturation of the rate for gluon emission by strong color field
effects (see also the discussion in Refs. \cite{AM02,PATH,GAUSS}).
By contrast, in the reaction--diffusion process, the particle
(occupation) number saturates at a {\em fixed} value (independent of
$Y$), a situation which looks unphysical from the perspective of
QCD: when increasing the rapidity even within the saturation domain,
the gluon phase--space opens towards lower longitudinal momenta,
which then allows for further radiation.

Closely related to the above, there is a second reason why the
Mueller--Salam model is better suited to mimic the QCD evolution
than the reaction model : unlike the latter, the former is
consistent with the {\em boost invariance} of the $S$--matrix in the
presence of multiple scattering. In fact, in Ref. \cite{AMSalam95},
this model has been constructed precisely by requiring that the
(average) $S$--matrix be independent upon the choice of a Lorentz
frame. Remarkably, within the zero--dimensional context, the
requirement of boost invariance together with the assumption of one
particle emission per unit rapidity are in fact sufficient to
uniquely fix the rate at which a system of $n$ particles can emit
another one, and predict, in particular, the saturation of this rate
for sufficiently large values of $n$. On the other hand, a model
including both particle splitting ($A \to AA$) and particle
recombination ($AA \to A$), is in conflict with boost invariance as
we shall explain in Appendix A.

In turn, these conceptual differences between the Mueller--Salam
model and the reaction--diffusion model entail important structural
differences between the two, as well as differences between their
respective predictions. For instance, we shall see that the
evolution described by the Mueller--Salam model involves
particle--number--changing vertices $m\to n$ for all values of $m$
and $n$, so like the JIMWLK equation and its recent generalizations
in Refs. \cite{KL05,KL3,BREM,Balit05}. By contrast, the reaction
model involves only $1\to 2$ and $2\to 1$ vertices. This difference
is in fact related to the issue of boost invariance in the presence
of multiple (eikonal) scattering: As recently shown in Refs.
\cite{KL3,BIIT05}, the requirement of boost invariance implies a
symmetry property for the evolution Hamiltonian known as {\em
self--duality}. As further explained in Ref. \cite{BIIT05}, on the
example of the `Pomeron loop' equations, the reaction--diffusion
dynamics can be made consistent with self--duality, but only under
the additional assumption that one can ignore multiple scattering
for the individual particles --- an assumption which looks unnatural
in the context of high--energy QCD. By contrast, if one allows for
multiple scattering in the eikonal approximation, then a self--dual
evolution Hamiltonian must involve an infinite number of
gluon--number--changing vertices, as explicit in the QCD
constructions in Refs. \cite{BREM,Balit05}.

Note also another feature of the reaction--diffusion models (at
least in zero transverse dimensions), which looks unphysical in the
context of QCD: As shown in Ref. \cite{KozLev06}, such models
predict a `grey disk' limit at high energy, that is, the
corresponding $S$--matrix saturates at a {\em non--zero} value when
$Y\to \infty$. This will be briefly discussed in Appendix A.

Other interesting results emerging from our study, may have a
counterpart in QCD as well. The nature of the physically relevant
configurations which contribute to the average $S$--matrix is {\em
frame--dependent}. For definiteness, consider the collision between
two systems which in their respective rest frames reduce to one
particle (the `target') and, respectively, to $n$ particles (the
`projectile'). While the average $S$--matrix $\langle S^n\rangle$ is
frame--independent, the configurations which dominate the average at
high energy are different in different frames: whereas in the rest
frame of the projectile, the average is dominated by {\em rare}
target configurations which are {\em dilute}, in the target rest
frame, and for $n\ge 2$, $\langle S^n\rangle$ is rather controlled
by the {\em typical} configurations in the wavefunction of the
projectile, which are {\em at saturation}. We shall find that these
typical configurations at saturation follow a Poisson distribution.
The case $n=1$ (the symmetric scattering) is special\footnote{Our
result for $\lan S\ran$ will be the same as in the original analysis
in Ref. \cite{AMSalam95}, where however the precise nature of the
relevant configurations has not been investigated.}: $\langle
S\rangle$ is controlled by rare configurations, but marginally
sensitive to the saturated ones, in any frame, and has a different
{\em subleading} behaviour at high energy as compared to the case
where  $n\ge 2$.

The paper is organized as follows: Sect. \ref{sec-toy} will be
devoted to structural aspects of the Mueller--Salam model. In Sect.
\ref{SECT_MASTER}, we shall construct the master equation from the
requirement of boost invariance; then, in Sect. \ref{SECT_EQSS}, we
shall use this equation to derive a hierarchy of evolution equations
for the $S$--matrix elements $\lan S^n\ran$. In Sect.
\ref{SECT_HIGHY} we shall use the solution to the master equation to
investigate the high--energy behaviour of the model. We shall
successively consider the $S$--matrix elements $\lan S^n\ran$ (in
Sect. \ref{SECT_SM}), the bulk of the particle distribution (in
Sect. \ref{SECT_PN}), and the frame--dependence of the
configurations which control the average $S$--matrix (in Sect.
\ref{SECT_BOOST}). Finally, in Sect. \ref{SECT_QCD}, we shall
discuss the correspondence between the toy model and the
high--energy problem in QCD, with emphasis on the similarity between
the respective evolution equations in various limits.

\section{The toy model: Structural aspects}\label{sec-toy}

As explained in the Introduction, the toy model that we shall
consider is `zero--dimensional' in the sense that its only variable
is the rapidity interval $Y$ which controls the high--energy
evolution. The `physical' problem that we have in mind is that of
the scattering between two systems of particles, referred to as the
projectile and the target. Each system is characterized by a
probability distribution giving the number of particles it contains
at a given rapidity. During the collision of the two systems,
particles of the projectile and the target  undergo independent
collisions, characterized by a scattering amplitude $\tau$.  In
order to keep close contact with QCD, we shall use throughout a
QCD--inspired terminology and refer to the two colliding systems as
two ``onia'' and to the particles which compose these onia as
``dipoles''. This is suggestive since, in the dilute regime at
least, the evolution that we shall describe corresponds to the
zero--dimensional version of Mueller's dipole picture
\cite{AM94,AM95,Salam95,AMSalam95}.

The number of dipoles in each system depends upon its rapidity, and
therefore on the frame. We put  the  right mover (the `target') at
rapidity $Y-Y_0$ and the left mover (the `projectile') at rapidity
$-Y_0$ ($Y_0>0)$. We denote by $P_m^{\rmL}(Y_0)$ the probability to
find exactly $m$ dipoles inside the projectile at rapidity $Y_0$.
Similarly $P_n^{\rmR}(Y-Y_0)$ will be the corresponding distribution
for the target. Allowing the dipoles in each onia to scatter
independently with each other, the $S$--matrix for a configuration
with $m$ dipoles in the projectile and $n$ dipoles in the target
takes the factorized form $S = \sigma^{m n}$, where $\sigma = 1-
\tau$ is the $S$--matrix for the scattering between two elementary
dipoles and $\tau$ is the corresponding $T$-matrix. The most
interesting case in view of the comparison with QCD is the weak
coupling regime $\tau\ll 1$ (see Sect. \ref{SECT_QCD}).

The physical  $S$--matrix is obtained by averaging over all
the possible configurations in the two onia, with the respective
probability distributions:
  \beq\label{eq-Stot}
  \avg{S}_{Y} = \sum_{m,n=1}^{\infty}
  P_{m}^{\rmL}(Y_0) P_n^{\rmR}(Y-Y_0)\,
  \sigma^{m n}.
  \eeq
This expression for the S-matrix reflects the fundamental
factorization assumption which lies at the basis of our analysis.
This formula was used  in Ref. \cite{KL5}, while in their original
formulation,  Mueller and Salam \cite{AMSalam95} write, in their
analog of Eq.~(\ref{eq-Stot}), $\rme^{-\tau mn}$ in place of
$\sigma^{mn}$.

\subsection{Particle saturation from boost invariance}
\label{SECT_MASTER}

We shall assume that the two onia follow the same evolution law in
rapidity, that is, they obey the same microscopic dynamics. The
initial conditions at $Y=0$ are left arbitrary for the time being,
but they will be specified later, when this will be needed.
Then, as we show now, the rapidity evolution is uniquely fixed by
the following two constraints:

\vspace*{.2cm}
\begin{enumerate}
\item[\tt(i)]{ Lorentz (boost) invariance:
The total $S$--matrix should be independent of the choice of frame,
i.e. , of  $Y_0$, which implies
  \beq\label{eq-LorInv}
  \frac{\dif\avg{S}}{\dif Y_0} = 0.
  \eeq}
\item[\tt(ii)]{ The onium evolves through dipole emission, in such a
way that only one new dipole can be produced under a step $\dY$ in
rapidity. Thus, in a step $\dY$, a system of $n$ dipoles can turn
into a system of $n+1$ dipoles, with a probability $f_n \dY$, and
stay in its initial configuration with a probability $1-f_n \dY$. It
follows that the probability $P_n(Y)$ evolves according to the
following master equation:
  \beq\label{eq-dPdY}
  \frac{\dif P_n(Y)}{\dY} =
  f_{n-1}\,P_{n-1}(Y) -
  f_n\, P_n(Y),
  \eeq
where $f_n$ is a function of $n$ to be  determined shortly. In this
equation and throughout, $P_n$ refers generically to
either $P_n^{\rmR}$, or $P_n^{\rmL}$. Note that the evolution
 (\ref{eq-dPdY}) preserves  the total
probability: $\dif(\sum_n P_n(Y))/\dif Y$ =0.}

\end{enumerate}
\vspace*{.2cm}

As noticed in Ref. \cite{AMSalam95}, the two conditions above
determine the transition probabilities $f_n$ up to an overall
normalization factor (which is then fixed by $f_1$). Indeed, setting
the derivative of Eq.~(\ref{eq-Stot}) with respect to $Y_0$ equal to
zero and using Eq.~(\ref{eq-dPdY}) we arrive at
  \beq\label{eq-fntemp}
  \sum_{m,n} P_{m}^{\rmL}(Y_0) P_n^{\rmR}(Y-Y_0)\,
  \sigma^{m n}
  [f_m (1-\sigma^n) - f_n (1-\sigma^m)] = 0
  \,\Rightarrow\,
  f_n = c\, (1-\sigma^n),
  \eeq
where $c$ is a constant that can be reexpressed as
$c=f_1/(1-\sigma)$, where $f_1$ is the rate of splitting of one
dipole into two. Without loss of generality, we shall set $f_1=1$.
(Within the QCD dipole picture, this rate would be equal to
$\abar/2\pi$ times the dipole kernel; see, e.g., Refs.
\cite{AM94,IM031} for details.)  We then obtain, from
Eq.~(\ref{eq-fntemp}),
  \beq\label{eq-fn}
  f_n = \frac{1-\sigma^n}{1-\sigma} =
  1+ \sigma +\cdots +\sigma^{n-1}
  \,\leq\, n.
  \eeq
The quantity $f_n$ is the rate at which a system of $n$ dipoles
emits an extra one in a step $\dY$ in rapidity. The constraint of
boost-invariance of the scattering matrix forces the  $f_n$ (and
hence the probability distribution $P_n$) to depend on $\sigma$, and
furthermore induces non trivial correlations that affect their
$n$--dependence. To better visualize the implications of
Eq.~(\ref{eq-fn}), it is useful to consider the two limits of strong
($\tau\to 1$) and weak ($\tau\to 0$) couplings. In the weak coupling
regime, $\sigma\simeq 1$ and $f_n\simeq n$: in this regime, the $n$
dipoles split independently of each other. In the strong coupling
regime however, $\sigma\simeq 0$, and $f_n\simeq 1$: in this case
correlations in the $n$-dipole system are such that only a single
dipole can be emitted.

When $\tau$ is fixed and small, which is the case of physical
interest, a similar change of regime occurs when increasing the
value $n$. Namely, Eq.~(\ref{eq-fn}) implies that, when $\tau\ll 1$,
the rate $f_n$ has the following limiting behaviors:
  \beq\label{eq-fnn}
    \frac{f_n}{n} =
    \begin{cases}
        \displaystyle{1-\frac{\tau (n-1)}{2}\,+\,\cdots} &
        \text{ for\,\,  $n \ll 1/\tau$}
        \\
        \displaystyle{\frac{1}{\tau n}\, -\, \cdots \ll 1} &
        \text{ for\,\,  $n \gg 1/\tau$}\,.
    \end{cases}
  \eeq
 According to Eq.~(\ref{eq-fnn}), the emission
rate $f_n$ saturates at a large, but constant, value $1/ \tau \gg 1$
when $n\simge n_{\rm sat}\sim 1/ \tau \gg 1$. As we shall discover
soon, $n_{\rm sat}$ is indeed the order of magnitude of the average
dipole number at the onset of saturation.

Alternatively, observing that $\sigma^n$ is the $S$--matrix for the
scattering of a dipole projectile on an $n$--dipole target,  we may
interpret the presence of higher powers of $\sigma$ in the emission
rate (\ref{eq-fn}) as reflecting  the {\em multiple scattering}
between the newly emitted dipole and its sources. This
interpretation becomes perhaps more transparent when
Eqs.~(\ref{eq-dPdY}) and (\ref{eq-fn}) are used to compute the rate
for the evolution of the average $S$--matrix with $Y$. One finds
 \beq\label{eq-dSdY}
  \frac{\dif\avg{S}}{\dif Y} = -
\frac{1}{\tau}\sum_{m,n} P_{m}^{\rmL}(Y_0) P_n^{\rmR}(Y-Y_0)\,
  \sigma^{m n}(1-\sigma^m)(1-\sigma^n),
  \eeq
where the quantities $t_n\equiv 1-\sigma^n$ and $t_m\equiv
1-\sigma^m$ can be recognized as the scattering amplitudes for the
scattering between one dipole --- the one created in the last step
in the evolution --- and the $n$ preexisting dipoles in the target
and, respectively, the $m$ dipoles in the projectile.

For latter reference, let us notice that Eq.~(\ref{eq-Stot}) is
equivalent to the following factorized form of the {\em scattering
amplitude} $\avg{T} \equiv 1-\avg{S}$
  \beq\label{eq-T}
  \avg{T}_Y = \sum_{k=1}^{\infty} \frac{(-1)^{k-1}}{k!}
  \lan n^{(k)} \ran_{Y_0}\,
  \lan t^k\ran_{Y-Y_0},
  \eeq
which is the form generally used in  the QCD context  (see, e.g.,
Refs. \cite{K,LL04,IST05}). In this equation,  $\lan n^{(k)} \ran
\equiv \lan n(n-1)\cdots(n-k+1)\ran$ is the $k$--body
normal--ordered dipole number (here, in the wavefunction of the
projectile), $t\equiv 1-s$ the scattering amplitude for a single
dipole, and $\lan t^k\ran_{Y-Y_0}$ is the average amplitude for the
simultaneous scattering of $k$ dipoles off the target. Note that, in
the discussion above, we have introduced the notation $s$ for the
$S$--matrix for a projectile made with a single dipole. This
notation will be systematically used in what follows.
Correspondingly, the $S$--matrix for a projectile made with exactly
$m$ dipoles is $s^m$.

Finally, let us mention that the $P_n$'s can be obtained from the
following generating functional
  \beq\label{eq-Z}
   Z(u,Y) = \sum_{n=1}^{\infty} P_n(Y)\, u^n.
  \eeq
from which most quantities of interest can also be derived. In
particular,
 \beq\label{eq-nkZ}
  \lan n^{(k)} \ran \equiv \lan n(n-1)\cdots(n-k+1)\ran
 \,= \,\frac{\dif^k Z(u,Y)}{\dif u^k}
    \Big|_{u=1}\,.
  \eeq
Differentiating Eq.~(\ref{eq-Z}) with respect to $Y$ and using the
master equation one arrives at
  \beq\label{eq-dZdY}
   \frac{\dif Z}{\dif Y} = \frac{1-u}{1-\sigma}\,
   \left[ Z(\sigma u) - Z(u) \right].
  \eeq
This equation turns out to be difficult to solve analytically, so in
what follows we shall work mostly with the master equation.

\subsection{Evolution equations: Saturation, unitarity \& fluctuations}
\label{SECT_EQSS}

Using the master equation (\ref{eq-dPdY}), we shall now deduce
evolution equations for physical quantities such as the average
$S$--matrix  or the average dipole number, and   study some of their
general properties. More specific predictions about the solutions to
these equations at high energy will be  discussed in Sect.
\ref{SECT_HIGHY}.

Let us begin with the scattering matrix and rewrite
Eq.~(\ref{eq-Stot}) as
 \beq\label{eq-Stot1}
  \avg{S}_{Y} = \sum_{m}P_m^{\rmL}(Y_0)\,
  \avg{s^m}_{Y-Y_0}
  \eeq
where
 \beq\label{eq-Sm}
 \avg{s^m}_Y\equiv \sum_{n} P_n^{\rmR}(Y)\,\sigma^{mn}\,=\,
 Z^{\rmR}(u=\sigma^m,Y)
  \eeq
is the average $S$--matrix for a projectile made with exactly $m$
dipoles and a generic target. Matrix elements like $\avg{s^m}_Y$
carry information about the dipole correlations in the target.

From Eqs.~(\ref{eq-Sm}), (\ref{eq-dPdY}) and (\ref{eq-fn}), it is
now straightforward to obtain the following evolution equation for
$\avg{s^m}$ :
 \beq\label{eq-dsn}
   \frac{\dif \avg{s^m}}{\dY} =f_m
   \left[\lan s^{m+1} \ran -
   \lan s^m \ran \right].
  \eeq
This is not a closed equation --- $\avg{s^m}$ being related to $\lan
s^{m+1} \ran$ ---, but rather a particular equation from an infinite
hierarchy. This equation has been obtained here by following the
evolution of the target (namely, by using Eq.~(\ref{eq-dPdY}) for
$P_n^\rmR$), but it can be easily reinterpreted as describing
evolution in the projectile: when increasing the projectile rapidity
by $\dY$, the incoming system of $m$ dipoles can turn, with a rate
$f_m$ into a system of $(m+1)$ dipoles, which then scatters off the
target with an $S$--matrix $\lan s^{m+1} \ran$. This is the origin
of the first term within the brackets in the r.h.s. of
Eq.~(\ref{eq-dsn}). As for the second term, involving
$(-\avg{s^m})$, it corresponds to the case where the system remains
intact during the evolution (which occurs with probability $1-f_m
\dY$).

For $m=1$ (a single dipole in the  projectile),
Eq.~(\ref{eq-dsn}) reduces to
  \beq\label{eq-ds1}
   \frac{\dif \avg{s}}{\dY} = \lan s^2 \ran -
   \lan s \ran,
  \eeq
which is formally identical to the first equation in the
Balitsky--JIMWLK hierarchy. However, differences with respect to the
latter appear already for $m=2\,$. Indeed we have
  \beq\label{eq-ds2}
   \frac{\dif \avg{s^2}}{\dY} = (2-\tau)
   \left[\lan s^3 \ran - \lan s^2 \ran \right].
  \eeq
The corresponding Balitsky equation\footnote{In the remaining part
of this section, by the ``Balitsky equations''  we shall mean the
simplified version of these equations at large--$N_c$ and for zero
transverse dimensions, that is, Eq.~(\ref{eq-dsn}) with $f_m\to m$.}
would not contain the term proportional to $\tau$  in the r.h.s.
More generally the Balitsky equations are obtained by replacing
$f_m$ by $ m$ in the r.h.s. of Eq.~(\ref{eq-dsn}), which amounts to
ignoring {\em saturation} effects in the  {\em projectile}. Such
effects may be expected to be negligible when $m\tau\ll 1$, but we
shall discover in the next section that this is not so. In fact they
are essential to get the correct description of the evolution at
large $Y$.

Alternatively, the difference between the hierarchy in
Eq.~(\ref{eq-dsn}) and the Balitsky hierarchy can be attributed to
{\em particle--number fluctuations} in the {\em target}. For $n\gg
1$ the discrete nature of $n$ becomes inessential and the associated
fluctuations are unimportant. Therefore, let us replace the
summation over $n$ in Eq.~{(\ref{eq-Stot})} for the average
$S$-matrix by the corresponding integration (but keep a discrete sum
over $m$ on the side of the projectile) and require boost
invariance. Then we find that we need to impose a separate evolution
law for the target and the projectile wavefunctions, that is, the
splitting rates $f_n^{\rmR}$ and $f_m^{\rmL}$ must be {\em
different} functions. It is a matter of simple algebra to
obtain\footnote{In Sect. \ref{SECT_QCD}, we shall argue that
Eq.~(\ref{eq-dPdYMF}) is the toy--model analog of the JIMWLK
equation.}
 \beq\label{eq-dPdYMF}
  \frac{\dif P_n^{\rmR}(Y)}{\dY}\, \simeq
  -\frac {\del}{\del n}\,
  \left[\frac{1-\sigma^n}{|\ln \sigma|}\,
  P_n^{\rmR}(Y)\right]
  \equiv
  -\frac {\del}{\del n}
  \left[f_n^{\rmR}\,
  P_n^{\rmR}(Y)\right]\,,
  \eeq
while at the same time the left mover is evolving according to
Eq.~{(\ref{eq-dPdY})}, but with $f_m^{\rmL} =m$. After also trading
the sum over $n$ in the definition (\ref{eq-Sm}) of $\avg{s^m}$ by
the corresponding integral, we find that Eq.~(\ref{eq-dPdYMF}) leads
to the Balitsky hierarchy, as anticipated:
 \beq\label{eq-dsB}
   \frac{\dif \avg{s^m}}{\dY}  \,\simeq\,
   m\,
   \left[\lan s^{m+1} \ran -
   \lan s^m \ran \right]\,.
  \eeq

The fact that neglecting the correlations associated with
particle--number fluctuations in the target  is equivalent to
ignoring the saturation effects in the projectile  is in agreement
with the general arguments in Ref. \cite{IT04}. Further insight on
this issue can be gained by rewriting the evolution equations in
terms of the scattering amplitudes $\lan t^k \ran$ introduced in
Eq.~(\ref{eq-T}). The corresponding equations are easily deduced
from those  for $\lan s^m \ran$ by using $t\equiv 1-s$. The first
three equations in this hierarchy read:
  \beq\label{eq-dt1}
   \frac{\dif \avg{t}}{\dY} &\,=\,& \lan t \ran -
   \lan t^2 \ran,\\
  \label{eq-dt2}
   \frac{\dif \avg{t^2}}{\dY} &\,=\,& 2 \lan t^2 \ran -
   2\lan t^3 \ran +
   \tau \lan t (1-t)^2 \ran,\\
  \label{eq-dt3}
   \frac{\dif \avg{t^3}}{\dY} &\,=\,& 3 \lan t^3 \ran -
   3 \lan t^4 \ran +
   3 \tau \lan t^2 (1-t^2) \ran
   +\tau^2 \lan t (1-t)^3\ran.
  \eeq
The new terms, as compared to the corresponding Balitsky equations,
are those proportional to $\tau$ or $\tau^2$ in the last two
equations. As mentioned before, these terms reflect dipole--number
fluctuations in the target.

First we consider Eq.~(\ref{eq-dt2}): Among the three fluctuation
terms there, namely $\tau \lan t (1-t)^2 \ran =\tau \lan t^3 -2t^2
+t \ran$, we shall focus on the last one, $\tau \lan t \ran$, since
this is the most important one\footnote{This is the analog of the
`fluctuation terms' which appear in the Pomeron loop equations
constructed in Refs. \cite{IT04,IT05}. See the discussion in Sect.
\ref{SECT_QCD}.}. (When $\tau\ll 1$, the other terms are negligible
in all regimes.) Although formally suppressed by a power of $\tau$
with respect to the (BFKL--like) term $\lan t^2 \ran$, the
fluctuation term $\tau \lan t \ran$ is in fact equally important in
the low density regime where $\lan t \ran \sim \tau$. To clarify its
physical meaning, note that, in the dilute regime, the average
dipole scattering amplitude is simply proportional to the average
dipole number in the target: $\lan t \ran \simeq \tau \avg{n}$.
Thus, $\tau \lan t \ran\simeq \tau^2 \lan n \ran$, and the physical
interpretation becomes transparent: under a rapidity step $\dY$, any
one among the $\avg{n}$ target dipoles can split into two and then
the child dipoles can scatter with the two projectile ones, with
strength $\tau^2$. Thus, the simultaneous scattering of two
projectile dipoles gives us access to the correlations induced via
dipole splitting in the target.

Let us now move to the next equations in the hierarchy.
Eq.~(\ref{eq-dt3}) for $\lan t^3 \ran$ contains two {\em relevant}
fluctuation terms, of order $\tau \lan t^2 \ran$ and $\tau^2 \lan t
\ran$, respectively. The first one has the same physical origin as
the term $\tau \lan t \ran$ in the equation for $\lan t^2 \ran$:
that is, two among the three projectile dipole `feel' a fluctuation
by scattering off the child dipoles produced by a splitting in the
target. The second term, $\tau^2 \lan t \ran\simeq \tau^3 \lan
n\ran$, describes the process in which the fluctuation is felt by
all the three projectile dipoles: e.g., two of them scatter off one
child dipole, and the third one scatters off the other child dipole.
These terms are both relevant since they are of the same order in
the low density region where $\avg{t} \sim \tau$. It is not hard to
understand the generalization to higher equations in this hierarchy:
the equation for $\lan t^k\ran$ will involve relevant fluctuation
terms of the following types: $\tau \lan t^{k-1} \ran$, $\tau^2 \lan
t^{k-2} \ran$, $\dots$, $\tau^{k-1} \lan t \ran$. All such terms
appear to be important for building up many--body correlations in
the dilute regime.

Incidentally, the previous discussion also shows that the r.h.s. of
the evolution equation for $\lan t^k\ran$ involves terms
proportional to $\lan t^m\ran$ where the power $m$ can take all the
values from $m=1$ up to $m=k+1$. Since $k$ is generic, this means
that the hierarchy as a whole involves vertices relating  $\lan
t^k\ran$ to $\lan t^m\ran$ for arbitrary values of $k$ and $m$.

Eq.~(\ref{eq-dt1}) allows us to estimate the rapidity $Y_c$ for the
onset of unitarity corrections in the dipole--target scattering. Let
us assume that, at $Y=0$, the target consists in a single dipole:
$P_n^{\rmR}(0)=\delta_{n1}$; then we have $\lan t^k\ran_0=\tau^k$,
hence $\lan t^{k+1}\ran_0 \ll \lan t^k\ran_0$ and these inequalities
will be preserved in the early stages of the evolution where one can
therefore neglect $\lan t^2\ran$ as compared to $\lan t\ran$ in the
r.h.s. of Eq.~(\ref{eq-dt1}). This then leads to $\lan t \ran\simeq
\tau \rme^{Y}$, the analog of the BFKL increase \cite{BFKL}. Clearly
this ceases to be correct when $\lan t \ran\sim 1$, that is, for
$Y\sim Y_c\equiv \ln(1/\tau)$. For larger rapidities $Y \simge Y_c$,
multiple scattering becomes important and ensures the unitarization
of the dipole amplitude ($\lan t \ran\to 1$ as $Y\to\infty$), as we
shall discover in the next section.

Equivalently, $Y_c$ marks the onset of the saturation effects in the
target in the frame where the projectile is dilute. To see this,
consider the evolution equation for the average dipole number
$\avg{n}$ in the target, that is
 \beq\label{eq-eqavn}
   \frac{\dif \avg{n}}{\dY} = \frac{1}{\tau}
 \left\lan 1 -\sigma^n\right\ran\,,
  \eeq
where $\lan 1- \sigma^n\ran=\lan t\ran$ is
recognized\footnote{Alternatively, the r.h.s. of Eq.~(\ref{eq-avgn})
is recognized as the average emission rate $\avg{f_n}$.} as the
scattering amplitude for a projectile made with one dipole, cf.
Eq.~(\ref{eq-Sm}). At low density, $\lan 1 -\sigma^n\ran \simeq \tau
\avg{n}$, and the dipole number exhibits the exponential growth
$\avg{n}=\rme^Y$ characteristic of BFKL evolution. But when
$\avg{n}$ becomes as large as $n_{\rm sat}\sim 1/\tau$, which
happens  for $Y\sim Y_c$, the growth is tamed by non--linear
effects. Eventually, when $Y\gg Y_c$,   $\lan \sigma^n\ran$ becomes
negligible so that ${\dif \lan n\ran}/{\dY} \simeq {1}/{\tau}$ and
$\lan n \ran$ grows only linearly (as expected for  the gluon
occupation factor in QCD \cite{AM99,SAT}). To summarize,
\beq\label{eq-avgn}
   \lan n \ran=\rme^Y\qquad\mbox{when}\qquad Y \ll Y_c\;;
   \qquad \lan n \ran \,\simeq\,
   \frac{Y - Y_c}{\tau}\qquad\mbox{when}\qquad Y \gg Y_c\,.
  \eeq

\section{The high--energy behaviour of the toy model}
\setcounter{equation}{0} \label{SECT_HIGHY}

In this section, we shall establish the dominant high--energy (i.e.,
large--$Y$) behaviour predicted by the toy model for the dipolar
$S$--matrix elements $\avg{s^m}$ with $m\ge 1$, and also for the
dipole distribution in the target.

\subsection{The dipole $S$--matrix elements $\avg{s^m}$}
\label{SECT_SM}

For definiteness, let us focus on the situation where the target
reduces to a single dipole in its own rest frame:
$P_n^{\rmR}(0)=\delta_{n1}$. (More general initial conditions will
be briefly discussed in Sect. \ref{SECT_BOOST}.) Also, let us
perform our calculation in the projectile rest frame. This last
choice turns out to be very non--trivial since, as we shall discover
in Sect. \ref{SECT_BOOST}, the configurations which dominate the
average $S$--matrix are different in different frames. In conformity
with these choices, the $S$--matrix element $\avg{s^m}$ will be
computed according to Eq.~(\ref{eq-Sm}); that is, we shall first
solve the master equation (\ref{eq-dPdY}) with the initial condition
$P_n^{\rmR}(0)=\delta_{n1}$ and then perform the sum over $n$ in
Eq.~(\ref{eq-Sm}). We shall not be able to perform these operations
exactly, but only under suitable approximations, which preserve the
dominant behaviour of $\avg{s^m}$ at high energy.

Before we proceed with our calculations, we note that the hierarchy
in Eq.~(\ref{eq-dsn}) admits the following, two--parameter, family
of solutions:
  \beq\label{eq-smexact}
  \lan s\ran_{\rm as} =
  \frac{A}{1-\sigma}(Y-Y_a)\, \rme^{-{Y}} \qquad\mbox{and}
  \qquad
   \lan s^m \ran_{\rm as} =
  \frac{A\sigma^{m-2}}{1-\sigma^{m-1}}\, \rme^{-{Y}}\quad\mbox{for}
 \quad m \ge 2\,,
  \eeq
with arbitrary values for the parameters $A$ and $Y_a$. Although
these cannot be the complete solutions, as they do not obey the
physical initial conditions at $Y=0$ for any choice of the free
parameters, with the choices $A=\sigma^2$ and $Y_a\sim Y_c \equiv
\ln(1/\tau)$ they describe the {\em asymptotic} form of the physical
solutions at large $Y\gg Y_c$ (hence the subscript `as'), as we
shall see.

What is remarkable about the behaviour of these solutions  is the
fact that, for asymptotically large $Y$, all the  $\avg{s^m}_Y$'s
approach the black disk limit ($S=0$) according to the {\em same}
exponential law $\exp({-Y})$,  for all $m$. This might look
counterintuitive since, at large $Y$, the target onium is typically
characterized by a large number of dipoles, off which a projectile
dipole will scatter with a very small $S$--matrix, $s\ll 1$. We
would then expect $\avg{s^m}\ll \lan s\ran$ when $m>1$. In fact, if
one assumes that $\avg{s^{m+1}}\ll \lan s^m\ran$ for sufficiently
large $Y$, then one gets from Eq.~(\ref{eq-dsn})
  \beq\label{eq-smMFA}
   \lan s^m \ran_{\rm typical} \approx \rme^{-f_m Y} \qquad\mbox{for \,
 large\, $Y$}\,,
  \eeq
which is indeed consistent\footnote{But, clearly, this consistency
disappears for large $m\simge 1/\tau$, which explains why the Ansatz
(\ref{eq-smMFA}) cannot be a solution to the complete hierarchy. In
fact, because of the coupling between small and large values of $m$
throughout the hierarchy, the estimate (\ref{eq-smMFA}) is wrong
even for relatively small $m\ge 2$, as clear from the comparison
with Eq.~(\ref{eq-smexact}).} with the initial assumption
$\avg{s^{m+1}}\ll \lan s^m\ran$ (at least, as long as $m\ll
1/\tau$), but is nevertheless in contradiction with the correct
asymptotic behaviour exhibited in Eq.~(\ref{eq-smexact}). The
problem with the estimate (\ref{eq-smMFA}) is that it implicitly
assumes that the average $S$--matrix is controlled by the {\em
typical} target configurations, which have a large dipole number.
However, even at large $Y$ there is still a non--vanishing
probability that dilute configurations remain present in the target,
and as we shall show explicitly,  the sum over $n$ in
Eq.~(\ref{eq-Sm}) is dominated by {\em rare fluctuations} for which
$n$ is relatively low, $n\sim\mathcal{O}(1)$, and therefore $s\sim
\sigma^{n}\sim 1$, which in turn implies $s^m\sim s$. Such
fluctuations have a low probability $\sim \exp({-Y})$, but this is
compensated by the fact that their contribution to the average
$S$--matrix is relatively large, of order one. On the other hand,
the typical configurations have a probability of order one. Since
$n$ is very large and of the order of the average dipole number
$\lan n \ran\simeq (Y-Y_c)/\tau$ (see Eq.~(\ref{eq-avgn})), the
contribution of these configurations to $\lan s^m\ran$ is
exponentially suppressed, namely $\sigma^{mn} \sim
\exp\{-m(Y-Y_c)\}$. These estimates suggest that indeed, at least
for $m\ge 2$, the average $S$--matrix $\lan s^m\ran$ is dominated by
the rare dilute configurations. The case $m=1$ is a priori more
subtle, since then both rare and typical configurations can give
significant contributions.

To make this discussion more quantitative and confirm the asymptotic
behaviour displayed in Eq.~(\ref{eq-smexact}), we shall now proceed
to an explicit calculation of the probabilities $P_n(Y)$. We do so
by using  the Laplace transform of  $P_n(Y)$
  \beq\label{eq-lappn}
   \wt{P}_n(\omega) =
   \int\limits_0^{\infty} \dif Y\,\rme^{-\omega Y} P_n(Y),
  \eeq
in terms of which the master equation reads
  \beq\label{eq-masterlap}
   \wt{P}_n(\omega) =
   \frac{f_{n-1} \wt{P}_{n-1}(\omega)+P_n(0)}{\omega +f_n}.
  \eeq
Using the initial condition  $P_n(0)= \delta_{n1}$, we get
  \beq\label{eq-lapsol}
   \wt{P}_n(\omega) = \frac{1}{f_n}
   \prod_{k=1}^n \left(1+\frac{\omega}{f_k}\right)^{-1},
  \eeq
from which one can obtain  $P_n(Y)$ by   the inverse Laplace
transform
  \beq\label{eq-Pny}
    P_n(Y) = \int\limits_{\mcal C} \frac{\dif \omega}{2 \pi \rmi}\,
    \rme^{\omega Y} \wt{P}_n(\omega),
  \eeq
where the integration is to be done in the counterclockwise
direction along any contour $\mcal{C}$ enclosing all the poles of
$\wt{P}_n(\omega)$. Notice that all the these poles occur in the
finite interval $(-1/\tau,-1]$.

There are two limiting behaviors of $P_n(Y)$ that can be identified
respectively with the cases of weak coupling ($f_n=n$) and strong
coupling ($f_n=1$). In the first case one obtains the familiar
distribution of the dipole model \cite{AM95}
  \beq\label{eq-dipoleY}
   P_n^{\rm dip}(Y) \,=\,\rme^{-Y}\,\big(1-\rme^{-Y}\big)^{n-1}\,.
  \eeq
  In the second case, the distribution is of the Poisson type:
  \beq\label{eq-poisson}
  P_n^{\rm Poisson}(Y)=\frac{Y^{n-1}}{(n-1)!}\,{\rm e}^{-Y}.
  \eeq
In the general case, we do not have a closed expression but one can
easily construct  $P_n(Y)$  in the form of a (finite) sum:
  \beq\label{eq-Pnseries}
   P_n(Y) = \sum_{k=1}^n c_k^n \,\rme^{-f_k Y},
  \eeq
with coefficients $c_k^n$  determined by
  \beq\label{eq-coeffite}
   \sum_{k=1}^n c_k^n = \delta_{n1} \quad \text{and} \quad
   c_k^n = \frac{f_{n-1}}{f_n-f_k}\,c_k^{n-1}.
  \eeq
These can be iteratively constructed. Here we shall present only
few of them, some of which will be important for our subsequent
analysis. We have
  \beq\label{eq-coeff}
   c_1^n = \frac{1}{\sigma^{n-1}}, \quad
   c_2^n = -\frac{f_{n-1}}{\sigma^{2n-3}},\quad \dots, \quad
   c_{n-1}^n = \frac{(-1)^n f_{n-1}}{\sigma^{n(n-1)/2}}, \quad
   c_n^n = \frac{(-1)^{n+1}}{\sigma^{n(n-1)/2}}.
  \eeq
Just for illustration, the first four probabilities are given by
  \beq\label{eq-P1}
   P_1(Y) &=& \rme^{-Y},\\ \label{eq-P2}
   P_2(Y) &=& \frac{1}{\sigma}\,\rme^{-Y} -
   \frac{1}{\sigma}\,\rme^{-f_2 Y},\\ \label{eq-P3}
   P_3(Y) &=& \frac{1}{\sigma^2}\,\rme^{-Y} -
   \frac{1-\sigma^2}{\sigma^3(1-\sigma)}\,\rme^{-f_2 Y}+
   \frac{1}{\sigma^3}\,\rme^{-f_3 Y},\\ \label{eq-P4}
   P_4(Y) &=& \frac{1}{\sigma^3}\,\rme^{-Y} -
   \frac{1-\sigma^3}{\sigma^5(1-\sigma)}\,\rme^{-f_2 Y}+
   \frac{1-\sigma^3}{\sigma^6(1-\sigma)}\,\rme^{-f_3 Y}-
   \frac{1}{\sigma^6}\,\rme^{-f_4 Y}.
  \eeq
  It is easy to verify that these distributions go over to either
  the dipole distribution (\ref{eq-dipoleY}) or the Poisson
  distribution (\ref{eq-poisson}) as $\sigma\to 1$ or $\sigma\to 0$,
  respectively.

\begin{figure}[t]
    \centerline{\epsfig{file=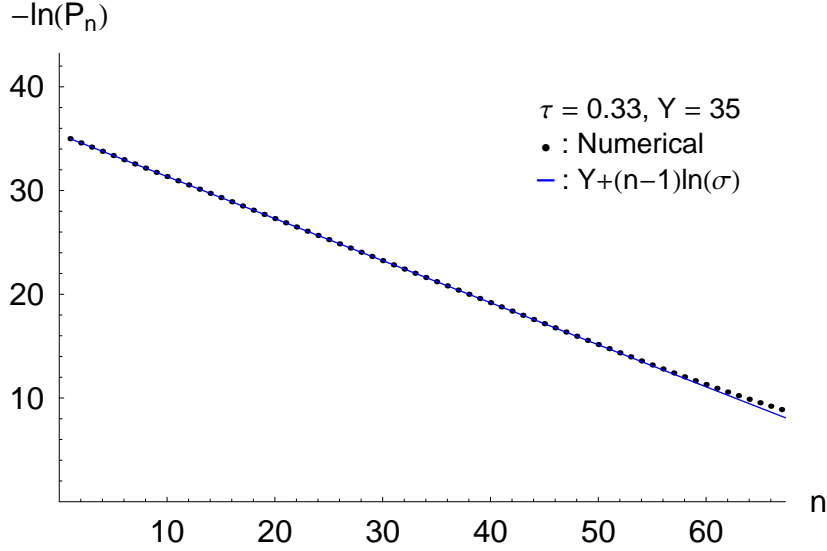,width=12cm}}
    \caption{\sl For high $Y$ and values of $n$
    such that $n \lesssim n_{\rm cr}$ the probabilities $P_n(Y)$
    are dominated by the first term of the sum in
    Eq.~(\ref{eq-Pnseries}). Notice that for the values of $Y$ and
    $\tau$ used in this plot, $n_{\rm cr}\sim 67$.}
    \label{figplow}
\end{figure}

From Eq.~(\ref{eq-Pnseries})--(\ref{eq-coeff}), it is clear that for
large $Y$ and not too large values of $n$, the dominant contribution
to $P_n(Y)$ is given by the first term in Eq.~(\ref{eq-Pnseries}),
proportional to $\rme^{-Y}$. Indeed, the terms with $k\ge 2$ are
exponentially suppressed at large $Y$ with respect to the first
term. But the situation changes when $n$ becomes  large,
since the coefficients $c_k^n$ increase rapidly with $n$ and this
rise can compensate for the exponential suppression with $Y$. One
can roughly estimate the value of $n$ at which this change of regime
occurs by requiring that the first two terms in the expansion of
$P_n(Y)$ become of the same order. This criterion yields
  \beq\label{eq-ncrit}
   n_{\rm cr} \sim \frac{\sigma Y - \ln(1/\tau)}{|\ln \sigma|}
   \sim \frac{Y-Y_c}{\tau},
  \eeq
where $Y_c\equiv\ln(1/\tau)$, as before, and the second estimate
holds when $\tau \ll 1$. A more precise estimate for $n_{\rm cr}$
will be obtained in the next section and reads $n_{\rm cr}\approx
(\sigma/\tau)(Y-Y_c)$. Clearly, this number is of the order of, but
smaller than, the average number of dipoles $\lan n\ran$ at large
$Y$ (cf. Eq.~(\ref{eq-avgn})). Hence, the configurations with $n\ll
n_{\rm cr}$ are relatively {\em dilute} and thus have a small
probability, approximately given by the first term in the sum in
Eq.~(\ref{eq-Pnseries}) : $P_n(Y)\approx
(1/\sigma)^{n-1}\rme^{-Y}\ll 1$. In Fig. \ref{figplow}, this
analytic estimate is compared with the exact result, as obtained by
the numerical solution to the master equation. On the other hand,
for $n\simge n_{\rm cr}$
--- the case of the {\em typical} configurations --- $P_n(Y)$ is of
order one and is dominated by the terms with large values of $k$, of
order $n$ (see Sect. \ref{SECT_PN}). As discussed above, we do not
expect these bulk configurations to affect the calculation of the
average $S$-matrix elements $\lan s^m \ran$, which are expected to
be controlled by the dilute configurations with $n\ll n_{\rm cr}$.

\begin{figure}[t]
    \centerline{\epsfig{file=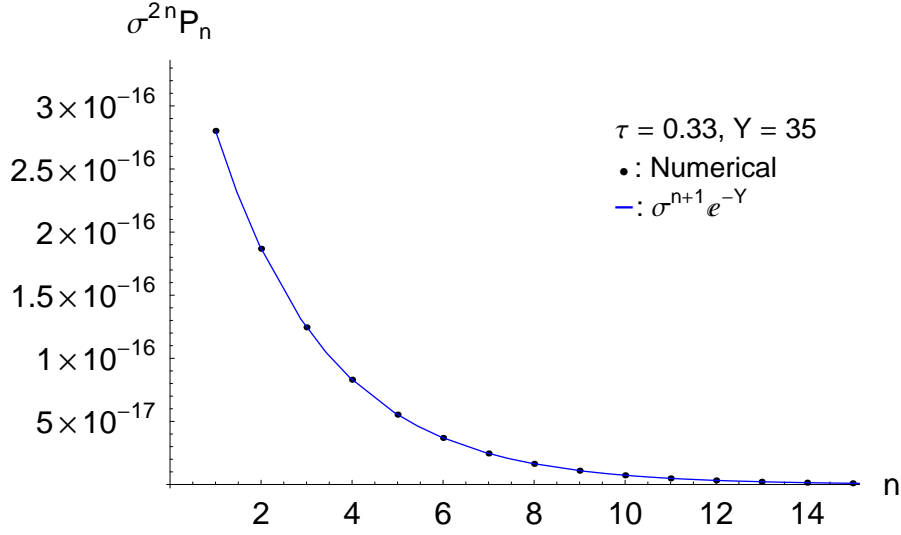,width=12cm}}
    \caption{\sl The distribution $\sigma^{2 n} P_n$ as a function of
    $n$ for large $Y$;
    only the rare configurations with $n$ up to $\sim 1/\tau$
    contribute to $\lan s^2 \ran$.}
    \label{figs2dist}\vspace*{2mm}
\end{figure}

Let us verify this  by explicitly computing $\lan s^m
\ran$, starting with $m=2$. To that aim, we separate out the
contribution of the configurations with $n \lesssim n_{\rm cr}$
in
Eq.~(\ref{eq-Sm}) for $\lan s^2 \ran$. This yields
  \beq\label{eq-s2sol}
   \lan s^2 \ran = \sum_{n=1}^{n_{\rm cr}} P_n(Y)\sigma^{2n}
   \simeq \rme^{-Y}\sum_{n=1}^{n_{\rm cr}}
   \frac{\sigma^{2n}}{\sigma^{n-1}} =
   \frac{\sigma^2 (1-\sigma^{n_{\rm cr}-1})}{1-\sigma}\,\rme^{-Y}
   \,\simeq\,
   \frac{\sigma^2}{1-\sigma}\,\rme^{-Y},
  \eeq
where the neglected terms are of  $\mcal{O}(\rme^{-2(Y-Y_c)})$, and
thus are exponentially suppressed when $Y\gg Y_c$ relatively to the
dominant contribution,  of $\mcal{O}(\rme^{-(Y-Y_c)})$. In practice
only the configurations with up to $\sim 1/\tau$ dipoles contribute
to the final result (this is manifest in Fig. \ref{figs2dist}), and
for them $\sigma^n \simge \sigma^{1/\tau} \sim \rme^{-1}$.  For any
such configuration, the $S$--matrix for a projectile dipole is of
order one, $s=\sigma^n\sim\mathcal{O}(1)$, as anticipated. The
contribution of the bulk configurations with $n \simge n_{\rm cr}$
to $\lan s^2 \ran$ will be considered in Sect. \ref{SECT_PN} and
found to be of $\mcal{O}(\rme^{-2(Y-Y_c)})$, so like the terms
neglected in evaluating Eq.~(\ref{eq-s2sol}). Hence, the final
result in Eq.~(\ref{eq-s2sol}) gives indeed the dominant behaviour
when $Y\gg Y_c$.

One can extend this calculation to an arbitrary $m \geq 2$. In fact,
the larger $m$ is, the faster the convergence of the sum over $n$ in
Eq.~(\ref{eq-Sm}) is\footnote{The extreme limit of this behaviour
occurs when $m$ becomes larger than $1/\tau$. Then the only
configuration of the target wavefunction which is relevant is the
one with $n=1$.}. Then it is straightforward to show that
  \beq\label{eq-smsol}
   \lan s^m \ran \simeq \frac{\sigma^m}{1-\sigma^{m-1}}\, \rme^{-Y}
  \qquad\mbox{for}
 \qquad m \ge 2\,.
  \eeq
We emphasize here that not only the $Y$--dependence, but also the
prefactor in the above equation are well under control. This result
is in agreement with the respective one in Eq.~(\ref{eq-smexact})
and it fixes the parameter $A$ there to be $A=\sigma^2$.

\begin{figure}[t]
    \centerline{\epsfig{file=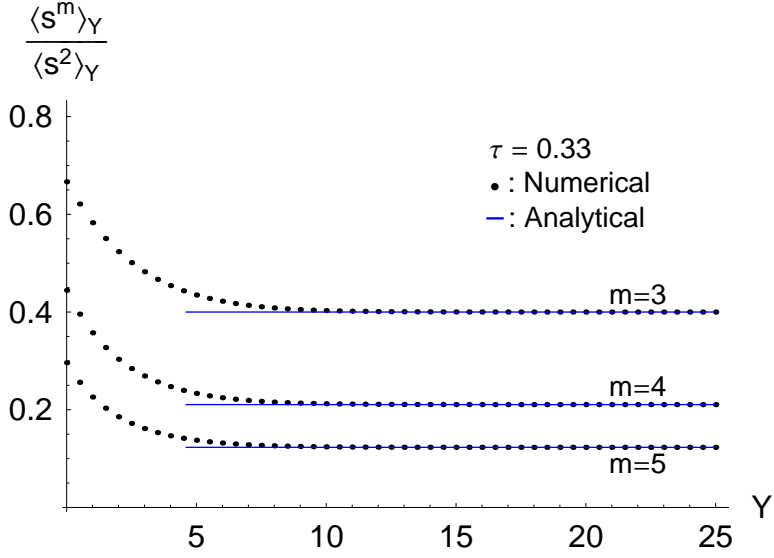,width=12cm}}
    \caption{\sl The $S$--matrix for the scattering of $m$
    projectile dipoles normalized to that for the
    scattering of two dipoles; for $Y \gg Y_c\approx 1.5$
    and $m \geq 2$, the $Y$--dependence is the same and equal to
    $\rme^{-Y}$.}
    \label{figratsmY} 
\end{figure}

\begin{figure}[t]
    \centerline{\epsfig{file=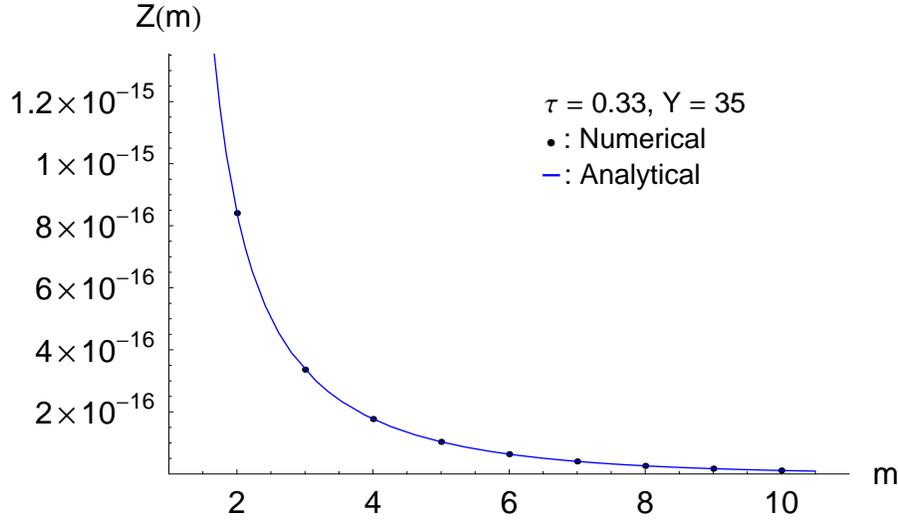,width=12cm}}
    \caption{\sl The generating functional as a function of
    $m = \ln u /\ln\sigma$ for $m>1$. When $m$ is an integer,
    $Z(m)$ gives the $S$-matrix for the scattering of $m$ projectile
    dipoles off the target.}
    \label{figZm}
\end{figure}

More generally, the above procedure allows one to calculate the
generating functional $Z(u,Y)$, Eq.~(\ref{eq-Z}), for large $Y$ and
any value of $u$ which is strictly smaller than $\sigma$: the
corresponding result is obtained by simply replacing $\sigma^m\to u$
in Eq.~(\ref{eq-smsol}), and reads
 \beq\label{eq-Zsmallu}
   Z(u,Y)\simeq \frac{u}{1- (u/\sigma)}\, \rme^{-Y}
  \qquad\mbox{for}\quad Y\gg Y_c \quad\mbox{and}\quad 0 \le u
 <\sigma\,.
 \eeq
The analytic estimate (\ref{eq-smsol}) for $\lan s^m \ran$ and the
corresponding one for $Z(m)\equiv Z(u= \sigma^m)$ are compared to
the respective numerical results in Figs. \ref{figratsmY} and
\ref{figZm}.

\begin{figure}[t]
    \centerline{\epsfig{file=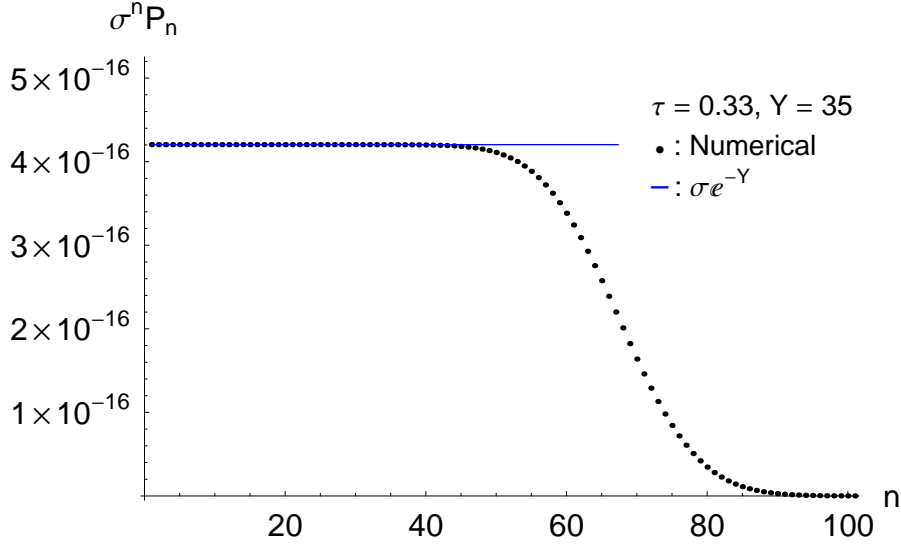,width=12cm}}
    \caption{\sl The distribution $\sigma^n P_n$ as a function of $n$
    for large $Y$;
    all the configurations with $n$ up to $n_{\rm cr}\approx
   \sigma (Y\!-\!Y_c)/\tau$
    contribute to $\lan s \ran$.}
    \label{figsdist}
\end{figure}

Let us now turn to the case $m=1$, which is special. Then the analog
of Eq.~(\ref{eq-s2sol}) reads
  \beq\label{eq-s2est}
   \lan s \ran = \sum_{n=1}^{n_{\rm cr}} P_n(Y)\sigma^{n}
   \simeq \sigma \rme^{-Y} \sum_{n=1}^{n_{\rm cr}} 1
   \,\simeq\,
   \frac{\sigma^2}{1-\sigma}\, (Y-Y_c)\,\rme^{-Y}\,,
  \eeq
where we have used the improved estimate  $n_{\rm cr}\approx
(\sigma/\tau)(Y-Y_c)$, to be found in  Sect. \ref{SECT_PN}. The
prefactor in front of the exponential in the final result is
essentially the number of configurations which contribute to the
average $S$--matrix, with each such configuration bringing a
contribution of $\mcal{O}(\rme^{-Y})$. Note, however, that the above
sum is dominated by its upper limit, i.e., by configurations with $n
\sim n_{\rm cr}$, for which our approximations are not fully under
control.  This result turns out to be nevertheless correct (up to
corrections of $\mcal{O}(1)$ to the rapidity shift $Y_c$, which go
beyond the present accuracy), for the following reason: the
distribution $\sigma^n P_n(Y)$ is almost flat as a function of $n$
so long as $n \lesssim n_{\rm cr}$
--- as manifest on Eq.~(\ref{eq-s2est}) --- but it drops out
very fast when $n> n_{\rm cr}$ (this  can be seen in the numerical
results in Fig. \ref{figsdist} and will be analytically verified in
the next section).

\begin{figure}[t]
    \centerline{\epsfig{file=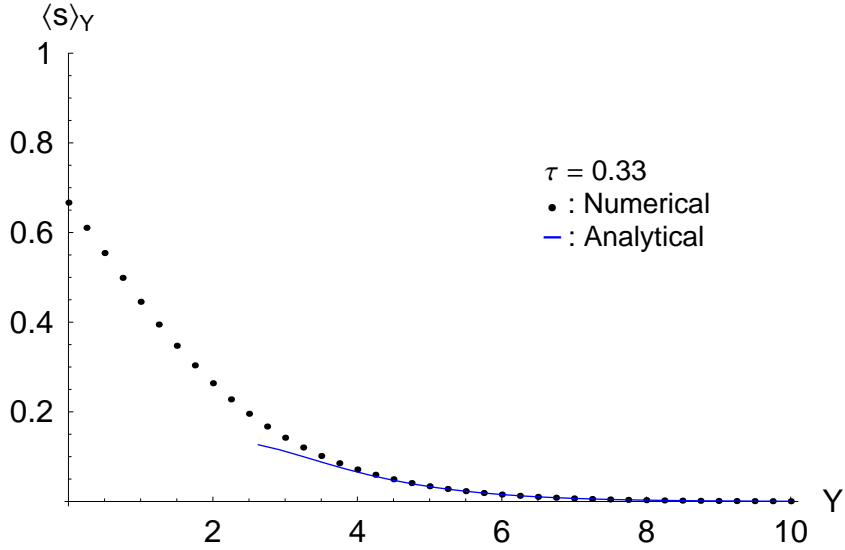,width=12cm}}
    \caption{\sl The expectation value $\lan s \ran$ of the
    $S$--matrix for onium--onium scattering. Starting from
    $\lan s \ran_{0} = \sigma$, this expectation values falls
    exponentially to zero when $Y \gg Y_c$,
    in agreement with the analytic prediction (\ref{eq-s2est})
    (see also Fig.~(\ref{figsYhigh})).}
    \label{figsY}
\end{figure}

\begin{figure}[t]
    \centerline{\epsfig{file=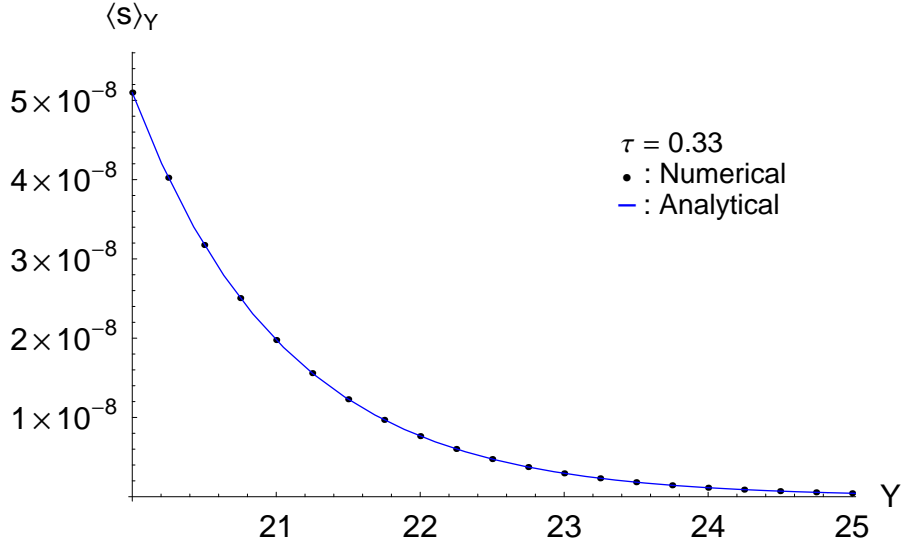,width=12cm}}
    \caption{\sl The average value $\lan s \ran$ for
    higher values of $Y$. The analytic curve is obtained by using
    Eq.~(\ref{eq-s2est}) with $Y_c=\ln(1/\tau)$.}
    \label{figsYhigh}
\end{figure}

A different way to check Eq.~(\ref{eq-s2est}) is to rely on the
hierarchy of evolution equations for  $\lan s^m \ran$. As previously
noticed, $\lan s^m \ran$ given by Eq.~(\ref{eq-smsol}) is an
accurate asymptotic solution for all the equations in the hierarchy
starting with the second one. Then one can use this result, together
with the first equation in the hierarchy, Eq.~(\ref{eq-ds1}), in
order to determine the asymptotic form of $\lan s \ran$. Thus one
easily recovers Eq.~(\ref{eq-s2est}) where however the rapidity
shift $Y_c$ is left undetermined. A further confirmation of
Eq.~(\ref{eq-s2est}) can be found in the original calculation of
$\lan s \ran$ by Mueller and Salam \cite{AMSalam95}, which is based
on the solution to Eq.~(\ref{eq-dPdYMF}) and which yields the same
result as our Eq.~(\ref{eq-s2est}) with $Y_c=
\ln(1/\tau)+\mcal{O}(1)$. The analytic estimate (\ref{eq-s2est}) is
compared to the respective, exact, numerical result in
Figs.~(\ref{figsY}) and (\ref{figsYhigh}), which confirm that both
the $Y$--dependence and the normalization shown in
Eq.~(\ref{eq-s2est}) are indeed under control.

To conclude this section, it is instructive to compare the above
results to those predicted by the dipole picture. Since the latter
is not boost invariant, we shall obtain different results for $\lan
s^m \ran$ depending upon the system that we decide to evolve: the
target or the projectile:

\texttt{(i)} {\em Projectile evolution in the dipole picture.} The
dipole $S$--matrix elements  $\lan s^m \ran$ obey the Balitsky
equations, that is, Eq.~(\ref{eq-dsn}) with the prefactor $f_m$
replaced by $m$. By doing this replacement, one looses Lorentz
invariance. This is formally recovered by allowing the projectile
and the target to obey different evolution equations. Namely, if the
projectile obeys dipole evolution, then in order for the r.h.s. of
Eq.~(\ref{eq-Stot1}) to be independent of $Y_0$, the target
probabilities $P_n^{\rmR}(Y-Y_0)$ which enter $\avg{s^m}_{Y-Y_0}$
via Eq.~(\ref{eq-Sm}) must evolve according to the `JIMWLK' version
of the master equation\footnote{This is, of course, in agreement
with the fact that the reduced Balitsky equations (\ref{eq-dsB}) can
be also obtained from the `JIMWLK' evolution (\ref{eq-dPdYMF}) of
the target.}, i.e., Eq.~(\ref{eq-dPdYMF}).

Let us then estimate the high--energy behaviour of $\lan s^m \ran$
as predicted by the Balitsky equations: by inserting the Ansatz
$\lan s^m \ran=\exp{(-c_m Y)}$ in this hierarchy, one immediately
finds $c_m=m$, and hence
  \beq\label{eq-smBAL}
   \lan s^m \ran = \rme^{-m Y} \qquad\mbox{ at large
 $Y$\ \ \ \ \ \ (no saturation in the projectile) }\,,
  \eeq
which is essentially the mean field estimate (\ref{eq-smMFA}). Thus,
by neglecting saturation effects in the projectile wavefunction (or,
equivalently, particle--number fluctuations in the target
wavefunction), we are led to the wrong conclusion that
$\avg{s^{m+1}}$ vanishes faster than $\lan s^m\ran$ at large $Y$.

\texttt{(ii)} {\em Target evolution in the dipole picture.} By using
the target--average expression (\ref{eq-Sm}) for $\lan s^m \ran$
together with the explicit solution (\ref{eq-dipoleY}) for the
probabilities in the dipole picture, one immediately finds:
 \beq\label{eq-smdip}
   \lan s^m \ran \simeq \frac{\sigma^m}{1-\sigma^m}\, \rme^{-Y}
  \qquad\mbox{for $m\ge 1$ \ and \ large
 $Y$}\qquad\mbox{(no saturation in the
 target)},
  \eeq
where the neglected terms are of order $\mcal{O}(\rme^{-2Y})$. For
$m\ge 2$, this is essentially the same as the correct result at
large $Y$, as given in Eq.~(\ref{eq-smsol}). This agreement is
consistent with the fact that $\lan s^m \ran$ with $m\ge 2$ is
dominated by rare target configurations with a small dipole number
and thus are insensitive to saturation effects. For $m=1$, on the
other hand, the dipole--picture prediction in Eq.~(\ref{eq-smdip})
is different from the correct respective result in
Eq.~(\ref{eq-s2est}); namely, it misses the large, overall, factor
$Y-Y_c$, which in Eq.~(\ref{eq-s2est}) has been produced by summing
over configurations with a relatively large number of dipoles
$n\simle (Y-Y_c)/\tau$ (cf. Eq.~(\ref{eq-s2est})), which are close
to saturation.

\subsection{The bulk of the dipole distribution}
\label{SECT_PN}

Although they appeared to be quasi--irrelevant in our previous
calculation of the average $S$--matrix, the typical configurations
become important when this calculation is performed in a different
frame. Therefore, in this subsection we study these typical
configurations, i.e. the probability distribution $P_n(Y)$ at high
energy (typically, $Y \gtrsim Y_c$) and for a large number of
dipoles $n \gg n_{\rm sat}$.

To compute this distribution, we return to the exact Laplace
transform, Eq.~(\ref{eq-lapsol}), and write
  \beq\label{eq-logPn}
   \ln[f_n \wtP_n(\omega)] =
   - \sum_{k=1}^{n} \ln\left(1+\frac{\omega}{f_k}\right)
   = \sum_{r=1}^{\infty} \frac{(-1)^r \omega^r}{r}
   \sum_{k=1}^{n} f_k^{-r},
  \eeq
where we have expanded the logarithm and exchanged the order of the
two summations. We shall now perform the summation over $k$ in the
above equation, under the assumptions that $n\tau \gg 1$ and $\tau
\ll 1$. More precisely, we shall neglect terms which are of order
(or smaller than) $\order{\rme^{-\tau n}}$ and/or $\order{\tau}$.
Notice that all these summations grow linearly with $n$ for large
$n$ (since $f_k^{-r} \simeq \tau^r$ when $k \gg 1/\tau$), so it will
be convenient to subtract this large contribution and then perform
approximations on the remainder. We thus have
  \beq\label{eq-sum1}
   \Sigma_1 \equiv \sum_{k=1}^{n} f_k^{-1}
   = \tau \sum_{k=1}^n \frac{1}{1-\sigma^k}
   \simeq \tau n + \tau
   \sum_{k=1}^\infty \frac{\sigma^k}{1-\sigma^k}
   \simeq \tau n + \ln(1/\tau) + \gamma_{\rm E},
  \eeq
where the third, approximate, equality has been obtained by
extending the upper limit of the sum from $n$ to $\infty$, which is
correct up to terms of order $\order{\rme^{-\tau n}}$. The ensuing
sum is evaluated in Appendix B under the assumption that $\tau\ll 1
$. Also, $\gamma_{\rm E} = 0.577...$ is the Euler constant.

Similarly for $r \geq 2$ we have (cf. Appendix B)
  \beq\label{eq-sumr}
   \Sigma_{r} \equiv \sum_{k=1}^{n} f_k^{-r}
   = \tau^r \sum_{k=1}^n \frac{1}{(1-\sigma^k)^r}
   \simeq \tau^r n + \tau^r
   \sum_{k=1}^\infty \frac{1- (1-\sigma^k)^r}{(1-\sigma^k)^r}
   \simeq \tau^r n + \zeta(r),
  \eeq
where $\zeta(r)$ is the Riemann Zeta function. Note that the above
summations are dominated by large values $k=\mcal{O}(n)$, as
anticipated in the discussion preceding Eq.~(\ref{eq-ncrit}). Using
Eqs.~(\ref{eq-sum1}) and (\ref{eq-sumr}), one can do the summation
over $k$ in Eq.~(\ref{eq-logPn}) to obtain
  \beq\label{eq-logPnsat}
   \ln[f_n \wtP_n(\omega)] \approx
   -\omega \ln(1/\tau)
   - n \ln(1 + \tau \omega)
   +\ln \Gamma(1+\omega).
  \eeq
Thus we finally arrive at
  \beq\label{eq-Pnsat}
   P_n(Y) \simeq  \tau\int\limits_{\mcal C} \frac{\dif \omega}
   {2 \pi \rmi}\,\frac{\rme^{\omega [Y - \ln(1/\tau)]}\,
   \Gamma(1+\omega)}
    {(1+\tau\omega)^n}\,,
  \eeq
which so far is valid for large $n\gg 1/\tau$ but arbitrary $Y$. It
is straightforward to check that this distribution satisfies indeed
the large--$n$ version of the master equation, i.e., the equation
obtained from Eq.~(\ref{eq-dPdY}) after replacing $f_n \simeq
1/\tau$, as appropriate when $n\gg 1/\tau$.

\begin{figure}[t]
    \centerline{\epsfig{file=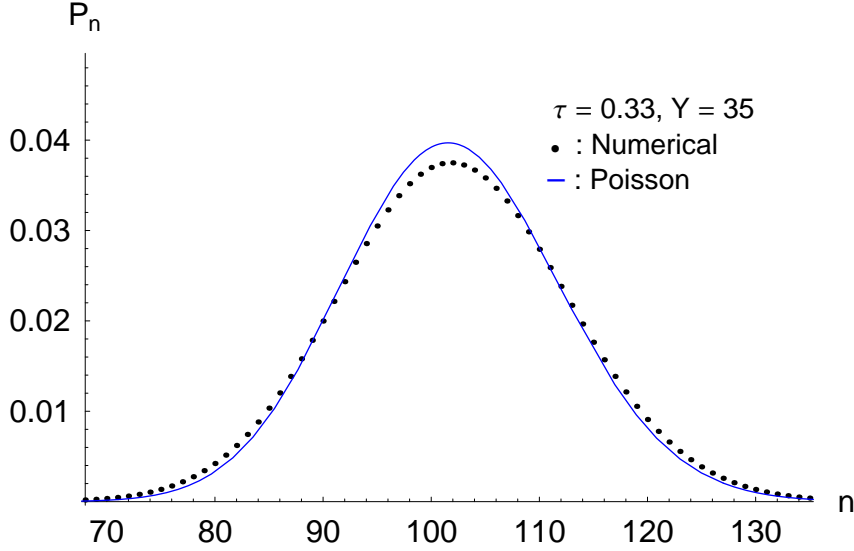,width=12cm}}
    \caption{\sl For high $Y$ the probabilities $P_n(Y)$
    follow a Poisson distribution for large values of $n$.}
    \label{figpoisson}\vspace*{.3cm}
\end{figure}

Even though Eq.~(\ref{eq-Pnsat}) is considerably simpler than the
general form, it is still difficult to proceed without any further
approximations, because of the presence of the Gamma function which
has single poles on the negative real axis. Since we are interested
in large values of $n$ and $Y$, we can evaluate the integral in
Eq.~(\ref{eq-Pnsat}) using a saddle point approximation. As we shall
see, the saddle point will occur at a small value of $\omega$, so
that we can replace the Gamma function by the lowest order terms in
its expansion around $\omega=0$ : $\Gamma(1+\omega)\simeq 1 -
\gamma_{\rm E} \omega$. (This means that only the sum $\Sigma_1$ in
Eq.~(\ref{eq-sum1}) and the $\tau^r n$ terms of the sums in
Eq.~(\ref{eq-sumr}) are kept in the subsequent analysis.) We then
obtain
  \beq\label{eq-Pnapprox}
   P_n(Y)\simeq
    \tau\int\limits_{\mcal C} \frac{\dif \omega}{2 \pi \rmi}\,
    \frac{\rme^{\tau\omega \nu}}
    {(1+\tau\omega)^n}
    \simeq  \frac{\tau}{\sqrt{2 \pi |F''(\omega_0)|}}
    \exp[F(\omega_0)],
  \eeq
where we have defined the variable
  \beq\label{eq-taudef}
   \nu = \frac{Y - \ln(1/\tau) -\gamma_{\rm E}}{\tau}\,,
  \eeq
and the function
  \beq\label{eq-F}
   F(\omega) = \tau \omega \nu -n \ln(1+ \tau \omega),
  \eeq
while the contour integral over $\omega$ has been evaluated in the
saddle point approximation. For the latter to be justified, at least
one of the conditions $\nu\tau \gg 1$ (i.e., $Y\gg  Y_c\equiv\ln
(1/\tau)$) and $n\tau\gg 1$ needs to be satisfied. Besides, the
saddle point must obey $|\omega_0| \ll 1$, for consistency with the
previous manipulations. Clearly, the saddle point occurs at
  \beq\label{eq-omega0}
   \omega_0 = \frac{n-\nu}{\tau \nu}\,,
  \eeq
so the condition $|\omega_0| \ll 1$ is tantamount to $|n-\nu| \ll
\tau \nu$. Evaluating $F$ and $F''$ at the saddle
point\footnote{Notice that the saddle point conditions require that
the integration contour crosses perpendicularly the real axis.} and
substituting in Eq.~(\ref{eq-Pnapprox}) we finally obtain a Poisson
distribution, as anticipated:
  \beq\label{eq-PnPoissontau}
   P_n(\nu)=
   \frac{1}{\Gamma(n)}\,\nu^{n-1} \rme^{-\nu}.
  \eeq
As aforementioned, this approximation is valid so long as $|n-\nu|
\ll \tau \nu$. This range covers indeed the ``bulk'' of the
distribution at large $Y$ (see also Fig.~\ref{figpoisson}) : when
summed over $n$ within this range, Eq.~(\ref{eq-PnPoissontau})
yields a total probability equal to 1 up to terms of order
$\order{\rme^{-\tau \nu}}$. Moreover, the lower limit $n_{\rm min}=
\nu - \tau \nu\equiv\sigma\nu$ is essentially the same as the upper
limit (previously denoted as $n_{\rm cr}$) of the validity range of
the approximation for $P_n$ constructed in Sect. \ref{SECT_SM}
(which, we recall, consists in preserving only the first term in the
sum in Eq.~(\ref{eq-Pnseries})). Indeed, if one redefines $n_{\rm
cr}$ via the condition that these two approximations match with each
other (up to prefactors) when $n=n_{\rm cr}$, then one finds $n_{\rm
cr}=(1-\tau +\order{\tau^2})\nu\approx n_{\rm min}$. Thus, the two
approximations for $P_n$ at large $Y$ that we have constructed in
this paper are complementary to each other, in the sense that,
together, they cover all the interesting values of $n$ and they
approximately match with each other at the borderline $n_{\rm
min}\equiv n_{\rm cr}$ between their respective domains of validity
in $n$.

Now let us discuss some aspects of this distribution. The maximum of
the distribution occurs at $n^*= \nu +1/2$ and the value at the
maximum is $P_{n^*}\equiv P_{\rm max} = 1/\sqrt{2 \pi \nu}$, while
the width of the distribution is proportional to $\sqrt{\nu}$. We
need to say here that terms of order $\order{\ln(1/\tau)}$  in the
location of the maximum have not been kept, since this would
requires a calculation of $\Sigma_1$ in Eq.~(\ref{eq-sum1}) to
$\order{\tau}$ accuracy (see Appendix B).

\begin{figure}[t]
    \centerline{\epsfig{file=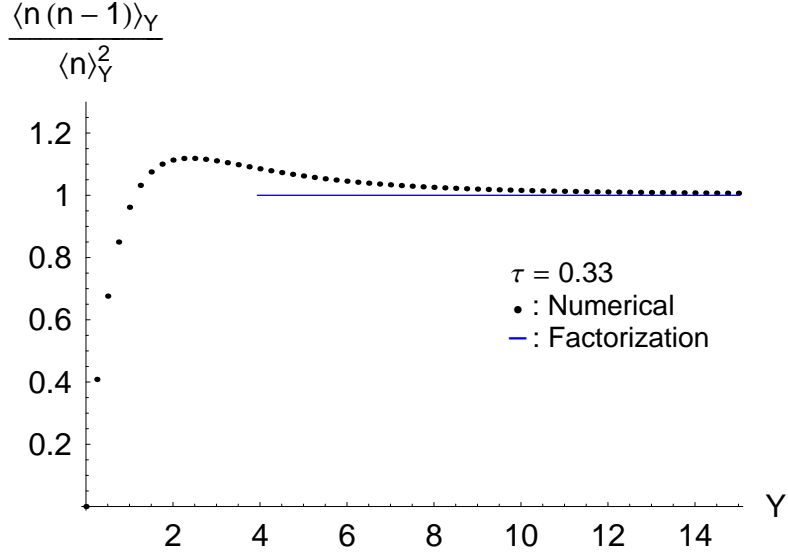,width=12cm}}
    \caption{\sl Factorization of expectation values holds only for
    quantities dominated by the bulk of the probability
    distribution; here the ratio of the (normal ordered) dipole-pair
    number with respect to the dipole number squared
    which approaches unity for $Y \gg Y_c$, in agreement
    with Eq.~(\ref{eq-avgnk}).}
    \label{figratnY}
\end{figure}

From Eq.~(\ref{eq-PnPoissontau}) is is straightforward to deduce the
average dipole number at high energy --- one thus finds $\lan n \ran
= \nu$, which is indeed the same as in Eq.~(\ref{eq-avgn})
--- and, more generally, all the $k$--body dipole
densities, which are readily obtained as\footnote{We should mention
here that the calculation of $\lan n^{(k)}\ran$ is consistent with
the validity range of the distribution (\ref{eq-PnPoissontau}) only
so long as $k \lesssim 1/\tau$.}
  \beq\label{eq-avgnk}
   \lan n^{(k)}\ran = \lan n \ran^k = \nu^k \qquad\mbox{when}\qquad Y
   \gg Y_c\,.
  \eeq
This equation exhibits mean-field--like factorization, as expected
for the bulk distribution at high density\footnote{Notice that this
is not the factorization that one finds in the dipole picture, i.e.,
in the absence of saturation; in that case one rather has $\lan
n^{(k)}\ran \simeq k! \lan n \ran^k$ at large $Y$ \cite{AM95}.}.
This is illustrated in Fig. \ref{figratnY}.

Let us now use the Poisson distribution (\ref{eq-PnPoissontau}) to
compute the contribution of the bulk configurations to the
expectation value of the $S$--matrix for the scattering of $m$
external dipoles. One thus finds
  \beq\label{eq-smbulk}
   \lan s^m \ran_{\rm bulk} =\,\sigma^m\, \rme^{-f_m \tau \nu}
    \sim\, \lan s \ran_{\rm bulk}^{f_m},
  \eeq
as expected from the mean field approximation, cf.
Eq.~(\ref{eq-smMFA}). However, for $m\ge 2$, this bulk contribution
is exponentially smaller (at large $\tau \nu \simeq Y-Y_c$) than the
respective contribution of the rare configurations with only few
dipoles, as previously computed in Eq.~(\ref{eq-smsol}). Also, even
for $m=1$, the result above is smaller by a large factor $1/(\tau
\nu)\ll 1$ than the previous result in Eq.~(\ref{eq-s2est}), which,
we remember, is due to the configurations with $n\simle n_{\rm cr}$.
Thus, the present calculation of the bulk distribution confirms our
previous conclusion that the $S$--matrix is dominated by the rare
configurations which involve relatively few dipoles.

The above results enable us to also compute the high--energy
behaviour of $Z(u,Y)$ for $u$ within the range $\sigma < u < 1$, and
thus complete our previous estimate, Eq.~(\ref{eq-Zsmallu}). Indeed,
for large $Y$ and $u>\sigma$, $Z(u,Y)$ is dominated by the typical
configurations, distributed according to
Eq.~(\ref{eq-PnPoissontau}). By using the latter within
Eq.~(\ref{eq-Z}), one easily finds
 \beq\label{eq-ZYlimit}
 Z(u,Y)\approx\exp\left\{- \frac{1-u}{\tau}(Y-Y_c)\right\}
 \qquad
 \mbox{for}\quad Y\gg Y_c \quad\mbox{and}\quad \sigma < u < 1.
 \eeq
The support of this function is strongly peaked near $u=1$, within a
distance $1-u\sim \tau/(Y-Y_c)$ which becomes smaller and smaller
with increasing energy and/or decreasing $\tau$. But if we let
$Y\to\infty$ for fixed $u< 1$ (even arbitrarily close to 1), then
$Z(u,Y)$ vanishes exponentially, which is another manifestation of
the `black disk' limit in our model.

\subsection{More on boost invariance and the initial conditions}
\label{SECT_BOOST}

Although the average $S$--matrix is frame--independent, by
construction, within the model under consideration, the physical
picture of the collision --- in the sense of relevant configurations
--- depends very much upon the choice of the frame, as we shall now
demonstrate via some explicit calculations. (This dependence has
been previously noticed in Refs. \cite{AMSalam95,IM032}.) To that
aim, it is convenient to consider the (toy--model analog of)
dipole--nucleus scattering; that is, we shall assume that, at $Y=0$,
the target is made with $A \ge 2$ dipoles, whereas the projectile
contains only one dipole: $P_n^\rmR(0) = \delta_{nA}$ and
$P_m^\rmL(0) = \delta_{m1}$. Note that this is the same physical
problem as considered in Sect.~\ref{SECT_SM} --- one onium starts as
a single dipole in its rest frame, while the other one starts as a
collection  of exactly $A$ dipoles (with $A$ denoted as $m$ in
Sect.~\ref{SECT_SM}) --- but the terminology that we use is now
different (in the sense of interchanging what we call `target' and
`projectile'), since we feel that the present terminology is more
natural in relation with a physical dipole--nucleus scattering.

For the  aforementioned initial conditions, we shall compute the
average $S$--matrix in two different frames: the rest frame of the
target (nucleus) and, respectively, that of the projectile (dipole).
For more clarity, we shall denote the results of these two
calculations by $\avg{S}_{1 \times A}$ and, respectively,
$\avg{S}_{A \times 1}$. More general initial conditions will be
briefly discussed towards the end.

\texttt{(a)} {\em Target (nucleus) rest frame}. When the nucleus is
at rest, the whole evolution is given to the dipole wavefunction.
This is precisely the situation analyzed in Sect.~\ref{SECT_SM},
from which we simply quote here the final result (cf.
Eq.~(\ref{eq-smsol}) with $m \to A$) :
 \beq\label{eq-s1A}
 \avg{S}_{1 \times A} =
 \frac{\sigma^A}{1-\sigma^{A-1}}\, \rme^{-Y}
 \qquad\mbox{for}
 \qquad A \ge 2
 \qquad (\text{nucleus rest frame})\,.
 \eeq
As also discussed in Sect.~\ref{SECT_SM}, in this frame the average
$S$--matrix is dominated by {\em rare} configurations in the evolved
system (the `dipole').

\texttt{(b)} {\em Projectile (dipole) rest frame}. This situation is
new with respect to Sect.~\ref{SECT_SM}, in the sense that the whole
evolution is now given to a system which starts with more than one
dipole at $Y=0$ (the `nucleus'). For simplicity we shall consider
first the case $A=2$ and then return to generic values for $A$. From
the arguments in Sect.~\ref{SECT_SM}, one may expect the average
$S$--matrix to be controlled by the rare configurations in the
evolved nucleus which are dilute (but this expectation turns out to
be wrong !), so let us first compute the respective contribution to
$\avg{S}$. Following the steps given in Sect.~\ref{SECT_SM}, but now
with a different initial condition, it is straightforward to find
that the distribution of the rare configurations reads
  \beq\label{eq-Pnrare2}
  P_n^{A=2}(Y) \simeq
  \frac{f_{n-1}}{\sigma^{2 n - 4}}\,
  \rme^{-f_2 Y}
  \qquad
  \text{for}
  \qquad 2 \le n \lesssim n_{\rm cr}\,,
  \eeq
where $n_{\rm cr}$ is basically the same as before, i.e., $n_{\rm
cr}\approx (\sigma/\tau)(Y-Y_c)$. (Note, however, that the value of
$Y_c$ depends upon $A$; see the discussion around Eq.~(\ref{eq-YcA})
below.) At a first glance, this result seems to imply that the
average $S$--matrix will be proportional to $\rme^{-f_2 Y}$, which
would be at variance with the previous result, Eq.~(\ref{eq-s1A}),
obtained in the nucleus rest frame and thus would signal a violation
of the boost invariance. However, this is not really the case, since
the summation over the dilute configurations is not convergent
anymore, rather it is dominated by its upper limit $n_{\rm cr}$
(which, we recall, is the borderline towards the bulk
configurations). Specifically,
  \beq\label{eq-sumsnPnrare}
   \avg{S}_{A \times 1} \sim
   \sum_{n=2}^{n_{\rm cr}}
   \sigma^n P_n^{A=2}(Y) \sim
   \rme^{-f_2 Y} \sum_{n=2}^{n_{\rm cr}} \sigma^{-n}
   \sim \,\rme^{-Y}\,,
  \eeq
where only the dominant exponential behaviour has been kept, and
this comes from the terms with $n\simeq n_{\rm cr}$, as anticipated.
This should be contrasted with the calculation in the nucleus rest
frame where the corresponding sum, cf. Eq.~(\ref{eq-s2sol}), is
rapidly convergent.

The above argument also shows that, in the present frame, $\avg{S}$
is rather dominated by the {\em typical} configurations in the
evolved nucleus. Returning to the case of arbitrary $A \ge 2$, and
following the procedure of Sect.~\ref{SECT_PN} with the modified
initial condition $P_n(0) = \delta_{nA}$, one finds that the precise
form of the Poisson distribution which describes the bulk is
  \beq\label{eq-bulkA}
   P_n^{A}(Y) =
   \frac{1}{(n-A)!}\,
   \nu^{n-A}\,
   \rme^{-\nu}\,,
  \eeq
with $\nu = [Y-Y_c(A)]/\tau$, and where the critical rapidity
$Y_c(A)$, determining the transition of the nucleus from the
unsaturated ``phase'' to the saturated one, reads (for $\tau \ll 1$)
  \beq\label{eq-YcA}
    Y_c(A) =
    \begin{cases}
        \displaystyle{\ln\frac{1}{A\tau}} &
        \text{ for\,\,  $A\tau \ll 1$}
        \\
        \displaystyle{0} &
        \text{ for\,\,  $A\tau \gg 1$}\,.
    \end{cases}
  \eeq
(Indeed, when $A\tau \gg 1$, the nucleus is saturated already at
$Y=0$. On the other hand, when $A\tau \ll 1$, the nucleus is dilute
in the early stages of the evolution, when the average dipole number
rises like $\lan n \ran= A\rme^Y$, cf. Eq.~(\ref{eq-avgn}), until it
reaches a saturation value of $\order{1/\tau}$; this happens when
$Y\sim Y_c$ with $Y_c$ as given above.) Given the probability
distribution Eq.~(\ref{eq-bulkA}), it is straightforward to
calculate the average $S$--matrix for the scattering of one dipole
off the bulk distribution of the nucleus. One finds
  \beq\label{eq-sA1}
   \avg{S}_{A \times 1} \,\approx\,
   \sigma^A\, \rme^{-[Y-Y_c(A)]} \,\approx\,
   \begin{cases}
        \displaystyle{\frac{\sigma^A}{A\tau}\,\rme^{-Y}} &
        \text{ for\,\,  $A\tau \ll 1$}
        \\*[0.3cm]
        \displaystyle{\sigma^A\,\rme^{-Y}} &
        \text{ for\,\,  $A\tau \gg 1$}
    \end{cases}
    \qquad (\text{dipole rest frame})\,.
  \eeq
Thus is the final result in the dipole rest frame and is indeed in
agreement with the corresponding one in the nucleus rest frame, as
given in Eq.~(\ref{eq-s1A}), so long as $\tau \ll 1$ and $A \gg 1$.

Thus we see that, whether the (average) $S$--matrix is dominated by
rare configurations or the bulk distribution, really depends upon
the frame that one decides to view the process in. This is not
perhaps not so surprising, since it reflects the fact that the
wavefunction of an evolved hadron is not a frame independent
quantity.

Notice also that there is some loss of accuracy when working with
the bulk distribution rather than with the rare configurations:
although the results (\ref{eq-s1A}) and (\ref{eq-sA1}) are
consistent with each other within their validity ranges, the former
is {\em exact} at high energy (i.e., it holds for any values of
$\tau$ and $A$), whereas the latter is only approximate (in our
second calculation, based on the bulk distribution, we have not been
able to obtain the expression of the prefactor when $A\tau \sim 1$).

Now let us consider that the two systems are initially composed of
$A$ and $B$ dipoles respectively, i.e. $P_n^{\rmR}(0) = \delta_{nA}$
and $P_m^{\rmL}(0) = \delta_{mB}$, and let us momentarily assume
that $A<B$. From the previous discussion one understands that it is
advantageous to work in the rest frame of the left mover. Then one
can compute the asymptotic form of the average $S$-matrix, since it
is dominated by the rare configurations of the right mover
wavefunction, and one finds
  \beq\label{eq-sAB}
   \avg{S}_{A \times B} \simeq
   \begin{cases}
    \displaystyle{\frac{\sigma^{A B}}
    {\prod_{k=1}^A (1-\sigma^{B-k})}\, \rme^{-f_A Y}} &
        \text{ for \,\,  $A<B$}
        \\*[0.5cm]
    \displaystyle{\frac{\sigma^{A (A+1)}}
    {\prod_{k=1}^{A}(1-\sigma^{A+1-k})}\,Y\rme^{-f_A Y}} &
    \text{ for\,\,  $A=B$}\,.
    \end{cases}
  \eeq
The average $S$-matrix for the special case $A=B$, has been obtained
from the $A$-th equation of the hierarchy in Eq.~(\ref{eq-dsn}),
after $\avg{S}_{A \times (A+1)}$ has been determined. Notice that
Eq.~(\ref{eq-sAB}) is an exact solution to this hierarchy.

More generally, one could imagine that the initial wavefunctions are
determined by some smooth distributions in the number of dipoles
(rather than by a given number of dipoles). Then one can determine
the $S$-matrix, by averaging $\avg{S}_{A \times B}$ over the initial
conditions. So long as $P_1^\rmR(0)$ and $P_1^\rmL(0)$ are
non--zero, the asymptotic behaviour is simply given by the
appropriate generalization of Eq.~(\ref{eq-s2est}), namely
 \beq\label{eq-sgen}
  \avg{S} =
  \sum_{A,B=1}^{\infty}
  P_A^{\rmR}(0)
  P_B^{\rmL}(0)
  \avg{S}_{A \times B}
  \simeq
  P_1^\rmR(0) P_1^\rmL(0)\,
  \frac{\sigma^2}{1-\sigma}\, Y \rme^{-Y}\,,
 \eeq
since the term $A=B=1$ dominates the double sum when $Y\to\infty$.

\section{Correspondence between the toy model and high energy QCD}
\setcounter{equation}{0}\label{SECT_QCD}

Throughout this paper, we have emphasized similarities between the
structure and the predictions of the toy model and the known or
expected properties of high--energy QCD. In what follows, we discuss
this correspondence in a more systematic way.

The factorization formula for the $S$--matrix, Eq.~(\ref{eq-Stot}),
is reminiscent of the factorization schemes proposed within
high--energy QCD in Refs. \cite{AM95,B1,IM031,KL3,BIIT05}, which
have in common to be symmetric between the projectile and the
target. With the noticeable exception of the dipole factorization
\cite{AM95}, which however fails to accommodate the saturation
effects in the wavefunctions of the colliding hadrons, all the other
schemes alluded to above are not written in terms of particle
numbers, but rather in terms of gluon fields, sometimes represented
(in the CGC formalism \cite{EdiCGC}) as classical color charges
together with the color fields they radiate. It is therefore
essential to establish the proper correspondence between gluons in
QCD and particles in the toy model.

This correspondence goes as follows: the `dipoles' in the toy model
correspond to $s$--channel gluons in QCD (so like the real color
dipoles in Mueller's dipole picture), while $\tau=1-\sigma$
corresponds to gluons exchanged in the $t$--channel. (The analog of
$\tau$ in QCD starts at order $\alpha_s^2$, corresponding to a
two--gluon exchange between a pair of dipoles.) Hence, in an
effective gluon language, the evolution described by
Eqs.~(\ref{eq-dPdY}) and (\ref{eq-fn}) contains vertices for gluon
splitting in the $s$--channel --- these are, of course, the vertices
$f_n$ --- and also vertices for both splitting and merging in the
$t$--channel --- each exchange $\tau$ being sandwiched between two
such vertices. Hence, this evolution constructs the analog of the
`Pomeron loops' in the onium wavefunction. Fully symmetric `Pomeron
loops' will develop, of course, in the $S$--matrix for onium--onium
scattering, as described by Eq.~(\ref{eq-Stot}).

With this identification, the toy--model hierarchy for the dipole
scattering amplitudes, cf. Eqs.~(\ref{eq-dsn}) or
(\ref{eq-dt1})--(\ref{eq-dt3}), corresponds to the QCD evolution
equations with Pomeron loops, as constructed in Refs.
\cite{IT04,IT05,MSW05}. To render this analogy more precise, let us
focus first on the case where Pomeron loops are absent, and for
which the toy model version is given in Eqs.~(\ref{eq-dPdYMF}) and
(\ref{eq-dsB}). We would like to argue that the continuum version of
the master equation, Eq.~(\ref{eq-dPdYMF}), is the toy--model analog
of the JIMWLK equation in QCD. The JIMWLK equation \cite{JKLW,W,CGC}
is written for the color fields $A^+_a\equiv \alpha^a$ radiated in
the $t$--channel by the gluons produced in the $s$--channel by the
high--energy evolution of the (right--moving) target. It is a
second--order, functional, differential equation with respect to
$\alpha$ and reads, schematically,
 \beq\label{JIMWLK}
 \frac{\del W_Y[\alpha]}{\del Y}\,=\,
 \frac{1}{2} \frac{\delta}{\delta \alpha_a}\,\chi^{ab}[\alpha]
 \frac{\delta}{\delta \alpha_b}\,W_Y[\alpha]\,,\eeq
where the transverse coordinates have not been shown. The quantity
$W_Y[\alpha]$ is the probability distribution for the fields
$\alpha_a$ (the analog of $P_n(Y)$ of the toy model). The kernel
$\chi^{ab}[\alpha]$ --- the analog of the emission rate $f_n$ --- is
non--linear in $\alpha$ to all orders, via Wilson lines. These
describe the multiple scattering of the gluon emitted in the
$s$--channel in the last step of the evolution off the background
field $\alpha$ (the $t$--channel gluons) produced by the previous
steps of the evolution; this rescattering is similar to that
included in $f_n$ within the toy model. To complete the
identification between Eqs.~(\ref{JIMWLK}) and (\ref{eq-dPdYMF}),
one should recall that in the JIMWLK equation each $s$--channel
gluon is allowed to radiate only two gluons in the $t$--channel.
Hence, the second--order derivative w.r.t. $\alpha$ in
Eq.~(\ref{JIMWLK}) can be interpreted as a single derivative w.r.t.
$n$ (the number of gluons in the $s$--channel) and then
Eq.~(\ref{JIMWLK}) becomes indeed similar to Eq.~(\ref{eq-dPdYMF}).

Strictly speaking, the JIMWLK equation {\em does} include some
fluctuations, since it is Fokker--Planck equation for a random walk
in the functional space of color fields\cite{PATH}. However, these
are merely {\em color} fluctuations which are suppressed at large
$N_c$. The essential correlations associated with {\em
gluon--number} fluctuations are lost because of the impossibility to
probe both ($s$--channel) gluons which are produced after one
splitting. This is related to the fact that, as aforementioned, an
$s$--channel gluon cannot radiate more than two $t$--channel fields,
and thus cannot undergo multiple scattering off the projectile. By
contrast, within the toy model, the dipoles {\em are} allowed to
scatter multiply
--- both inside the target wavefunction and with the dipoles in the
projectile ---, hence the $n$--body correlations associated with
splitting are fully taken into account. In the language of QCD, each
$s$--channel gluon is allowed to absorb and radiate an arbitrary
number of $t$--channel gluons, so like in the more general,
`self--dual', evolution described in Refs. \cite{BREM,Balit05}.

In fact, as noticed in Ref. \cite{KL5}, the structure of the master
equation for the toy model bears some formal resemblance to that of
the `diamond' Hamiltonian constructed in Refs. \cite{BREM,Balit05}.
In particular, it shares with the latter the property of being
self--dual, as it should, since the self--duality of the Hamiltonian
is equivalent to the fact that the $S$-matrix is boost--invariant,
when the latter is given by a factorization formula like the one in
Eq.~(\ref{eq-Stot}) \cite{KL3,BIIT05}. To see this, note that the
master equation can be rewritten as (cf. Eqs.~(\ref{eq-dPdY}) and
(\ref{eq-fn}))
  \beq\label{eq-dPdual}
  \frac{\dif P_n(Y)}{\dY} \,=\,
   -\,\frac{1}{\tau}\,\Big(1-\rme^{-\frac {\del }{\del n}}\Big)
   \Big(1-\rme^{-n|\ln \sigma|}\Big)
   \, P_n(Y)\,\equiv\,H\Big[n|\ln \sigma|,
 \frac {\del }{\del n}\Big]P_n(Y)\,,
  \eeq
where we have extended $n$ to be a continuum variable and used
$g(n-1) = \exp(-\dif/\dif n)g(n)$ for a generic function $g(n)$.
As anticipated, the `Hamiltonian' appearing in Eq.~(\ref{eq-dPdual})
is `self--dual' \cite{KL5}, that is, it is invariant under the
self--duality transformation which in the present context consists
in exchanging
 \beq\label{eq-selfdual}
 \frac {\del }{\del n}\ \longleftrightarrow \ n|\ln \sigma|\,,
 \eeq
and then reversing the order of the operators. Moreover, the
presence of two types of exponentials --- one involving the number
of particles $n$ and one involving the derivative ${\del }/{\del n}$
--- is reminiscent of the two types of Wilson lines which appear in
the QCD Hamiltonian proposed in Refs. \cite{BREM,Balit05}. In QCD,
$n|\ln \sigma|$ is replaced by the color field produced by the
$s$--channel gluons, and  ${\del }/{\del n}$ by the (functional)
derivative with respect to the color charge density of these gluons.

Let us now return to the equations obeyed by the dipole scattering
amplitudes and compare the structure of the fluctuation terms
between the toy model and QCD. Consider first the equation for $\lan
t^2 \ran$, as displayed in Eq.~(\ref{eq-dt2}) for the toy model and,
respectively, in Eqs.~(2.7) and (2.8) of Ref. \cite{IT05} for QCD.
In both cases, the dominant fluctuation term is of the generic type
$\tau\lan t \ran$, i.e., it is linear in $\lan t \ran$ and of order
$\tau\sim\alpha_s^2$. Yet, the experience with the toy model tells
us that the coefficient of the fluctuation term in QCD (cf.
Eq.~(2.7) in Ref. \cite{IT05}) may actually be incomplete. Indeed,
in the calculation of this term in Refs. \cite{IT04,IT05,MSW05} one
has neglected the possibility that the individual dipoles from the
target undergo multiple scattering with the dipoles in the
projectile; that is, in deriving the equation for $\lan t^2 \ran$,
the two external dipoles have been allowed to scatter only with two
{\em different} dipoles from the target (but not also to scatter
both off a {\em same} dipole). By contrast, in the toy model
calculation, multiple scattering {\em is} included and is in fact
responsible for an $\order{1}$ contribution to the coefficient of
fluctuation term $\tau \lan t \ran$ in Eq.~(\ref{eq-dt2}) : if that
contribution was neglected, the ensuing coefficient would be twice
as large than the correct one. More generally, if one considers the
equation satisfied by $\lan t^{m} \ran$ within the toy model, then
among the $m-1$ (relevant) fluctuation terms which are included in
this equation (cf. Sect. \ref{SECT_EQSS}) --- namely, $\tau \lan
t^{m-1} \ran$, $\tau^2 \lan t^{m-2} \ran$, $\dots$, $\tau^{m-1} \lan
t \ran$ --- only the analog of the first term, $\tau \lan t^{m-1}
\ran$, has been so far included in the corresponding equation in QCD
\cite{IT05}.

This discussion allows us to draw some lessons for QCD: most likely,
the Pomeron loop equations in QCD at large $N_c$ should be completed
by including the effects of the multiple scattering of the
individual target dipoles. This conclusion appears to be in conflict
with some recent analyses within QCD \cite{MMSW05,HIMS05}, which
should be therefore carefully reexamined. Furthermore, the more
intricate structure for the fluctuation terms suggested by the toy
model seems to prevent one from mapping the problem under study into
a single Langevin equation. Note that, although the detailed
structure of the fluctuation terms in QCD may be indeed more
complicated than anticipated by the original analysis in Refs.
\cite{IT04,IT05,MSW05,BIIT05}, it is possible that the additional
terms will not modify the qualitative asymptotic behaviour of the
dipole scattering amplitudes at high energy and large $N_c$. As
argued in Refs. \cite{IMM04,IT04}, this behaviour seems to be
dictated simply by the stochastic nature of the evolution equations.

An other interesting result of the toy model that we expect to
extend to QCD as well is the fact that the average $S$--matrix in
the vicinity of the unitarity limit is dominated by rare
fluctuations with only few gluons. This is in agreement with the
arguments presented in that sense in Refs. \cite{AMSalam95,IM032},
which also show that the exponential approach towards the black disk
limit should be somehow faster in QCD, namely like\footnote{This is
also the behaviour found by Salam, via numerical simulations of the
onium--onium scattering within the context of the dipole picture
\cite{Salam95,AMSalam95}.} $\lan s\ran \sim \exp(-Y^2)$, to be
compared with the toy--model result $\lan s\ran \sim \exp(-Y)$. (The
additional factor of $Y$ in QCD comes from the phase--space for
diffusion in the dipole transverse sizes.) But what is most
interesting about our present results is that the $S$--matrix $\lan
s^m\ran$ for the simultaneous scattering of {\em several} dipoles is
even more strongly dominated by the rare fluctuations which are
dilute, to the point that $\lan s^m\ran$ with $m\ge 2$ is not at all
sensitive to saturation effects in the target wavefunction, and
(unlike $\lan s\ran$) it could have been simply computed within the
dipole picture. It would be interesting to understand whether a
similar feature holds in QCD as well. If so, this would mean that,
for $m\ge 2$, $\lan s^m\ran$ in high--energy QCD at large $N_c$ can
be reliably computed via numerical simulations within Mueller's
dipole picture.

\vspace*{-3mm}
\section*{Acknowledgments}
\vspace*{-3mm}

We are grateful to Al Mueller for useful discussions on the
manuscript. Also, two of us (E.I. and D.T.) would like to thank
Misha Kozlov, Genya Levin and Arif Shoshi for patient explanations
concerning their respective works on toy models.

\vspace*{-2mm}

\appendix

\section{Lack of boost invariance for the recombination
process}\label{SECT_NOBOOST}
\setcounter{equation}{0}

In this appendix, we shall consider more general (zero--dimensional)
models in which in addition to particle splitting, one allows for
recombination. The simplest model of this type is the reaction model
$A \rightleftharpoons AA$, introduced in the context of QCD in Ref.
\cite{IT04}, and further discussed in Refs.
\cite{Stasto05,SX05,KozLev06}. In that model, a particle can split
into two ($A \to AA$) at a rate $\alpha$ per particle and,
conversely, two particles can recombine into one ($AA \to A$) with a
rate $\beta$ per pair of particles. In what follows, we shall
consider a more general version of this model, in which the rates
for splitting ($f_n$) and merging ($g_n$) are taken to be general
functions of the number $n$ of particles in the system. Hence, in
such a process,  particle number saturation can in principle occur
via two different mechanisms: the saturation of the rate for
particle splitting and the particle recombination.

However, as we shall show in what follows, the recombination process
cannot be made consistent with the boost invariance of the
$S$--matrix (within the factorization scheme of Eq.~(\ref{eq-Stot}))
for any choice of the $g_n$'s. This finding, together with the fact
that, within the context of the JIMWLK equation, gluon saturation
occurs via the saturation of the emission rate, suggests that the
models based on recombination may not closely resemble the QCD
dynamics. This conclusion is further supported by the observation in
Ref. \cite{IST05} that, if one attempts to interpret the Pomeron
loop equations of QCD \cite{IT04} as a reaction--diffusion process
(in either the target or the projectile), then one is lead to
introduce an effective `recombination vertex' which has no definite
sign, and hence no probabilistic interpretation.

The master equation is obtained by adding to the r.~h.~s.~of
Eq.~(\ref{eq-dPdY}) the terms responsible for recombination. We thus
obtain
  \beq\label{eq-dPdYrec}
  \frac{\dif P_n(Y)}{\dY} =
  f_{n-1}\,P_{n-1}(Y) -
  f_n\, P_n(Y)
  +g_{n+1}\,P_{n+1}(Y) -
  g_n\, P_n(Y),
  \eeq
where the functions $f_n$ and $g_n$ should be chosen so as to ensure
the boost invariance of the $S$--matrix and satisfy the boundary
conditions $f_1=\alpha$ and $g_1=0$. By requiring the average
$S$--matrix in Eq.~(\ref{eq-Stot}) to be independent of $Y_0$, we
easily obtain the following constraint
  \beq\label{eq-fgboost}
   f_m (1-\sigma^n) - f_n (1-\sigma^m)
   -g_m (\sigma^{-n}-1) + g_n (\sigma^{-m}-1) =0.
  \eeq
A priori, there are two different ways in which this constraint
could be satisfied: \texttt{(i)} the independent cancellation of the
$f$-terms and $g$-terms, and \texttt{(ii)} the mutual cancellation
between the two types of terms. Both are excluded, as we now show.

\texttt{(i)} By requiring the independent cancellation of the
$f$-terms and the $g$-terms in Eq.~(\ref{eq-fgboost}), one finds
  \beq\label{eq-fgind}
   f_n = c\, (1- \sigma^n) \quad  \text{and} \quad
   g_n = d\, (\sigma^{-n} - 1),
  \eeq
where the constants $c$ and $d$ are determined by the boundary
conditions at $n=1$. For $f_n$ we recover Eq.~(\ref{eq-fn}), but the
condition $g_1=0$ can be satisfied only by choosing $d=0$, and hence
$g_n=0$ for any $n$.

\texttt{(ii)} In order for the $f$-terms to cancel against the
$g$-terms in Eq.~(\ref{eq-fgboost}) one needs\footnote{From the
perspective of Eq.~(\ref{eq-Stot}), this would correspond to a
scenario in which the emission of a dipole in one onium can be
reinterpreted, after a shift $\dY$ in $Y_0$, as a recombination in
the other onium.}
  \beq\label{eq-fgdep}
   f_n = c\, (\sigma^{-n} - 1) \quad  \text{and} \quad
   g_n = -c\, (1-\sigma^{n}).
  \eeq
Depending on the sign of the constant $c$, either the $f_n$'s or the
$g_n$'s will be negative, and thus there is no probabilistic
interpretation --- a situation reminiscent of the difficulty met in
Ref. \cite{IST05}.

The standard reaction--diffusion process $A \rightleftharpoons AA$
is defined by (see, e.g., \cite{IT04})
  \beq\label{eq-reacdiff}
   f_n = \alpha n \quad  \text{and} \quad
   g_n = \beta n (n-1)/2,
  \eeq
and thus it does not belong to the case \texttt{(ii)}; this process
has a well--defined probabilistic interpretation, but is
inconsistent with the boost invariance of the $S$--matrix in the
factorization scheme of Eq.~(\ref{eq-Stot}), which includes multiple
scattering for the individual particles\footnote{In Ref.
\cite{BIIT05}, the Pomeron loop equations in QCD at large--$N_c$
have been given an effective interpretation in terms of a
reaction--diffusion process with `BFKL Pomerons', which appears to
be self--dual and thus formally consistent with boost invariance.
Note, however, that in that construction, one has given up multiple
scattering, that is, one has used a different factorization scheme
in which the individual dipoles (from the target and the projectile)
are allowed to scatter only once.}.

To better appreciate this difficulty, let us explicitly compute,
within the reaction model of Eq.~(\ref{eq-reacdiff}), the
high--energy limit of the average $S$--matrix for `dipole--nucleus
scattering' --- i.e., for the scattering between two onia which in
their respective rest frames reduce to a single dipole in the case
of the projectile and, respectively, to a set of $A$ dipoles, with
$A\ge 1$, in the case of the target --- by separately working in the
target rest frame and in the projectile rest frame. We anticipate
that the final results will be non--zero in both cases (in agreement
with the `grey disk limit' reported in Ref. \cite{KozLev06}), but
{\em different} for the two calculations. Only one of these results
will correspond to that found in Ref. \cite{KozLev06}, which has
adopted the point of view of the projectile evolution.

Before we proceed, let us notice that, within the reaction model of
Eq.~(\ref{eq-reacdiff}), the quantities $\alpha$, $\beta$ and
$\sigma$ are a priori independent parameters (the first two of them
referring to the evolution of the onium wavefunction, and the third
one characterizing the elementary dipole--dipole scattering), but in
view of the correspondence with QCD one needs to assume that the
ratio  $\beta/\alpha$ is of the same order as $\tau\equiv 1-\sigma$,
namely they are both of $\order{\alpha_s^2}$ (in particular, this
implies $\beta\ll \alpha$). To facilitate the comparison between
this model and the one studied in this paper, we denote
$\beta/2\alpha\equiv\tau_0$, with $\tau_0\sim\tau\ll 1$.

We start by summarizing the formul\ae\ giving $\avg{S}_{Y}$ in the
considered frames:

\texttt{(a)} {\em Target (nucleus) rest frame} (i.e., projectile
evolution). This is the case $Y_0=Y$ in Eq.~(\ref{eq-Stot}), which
then implies:
  \beq\label{eq-STRF}
  \avg{S}_{Y}^{\rmL} = \sum_{m}P_m^{\rmL}(Y)\,
  \,\sigma^{mA}\,=\,Z^{\rmL}(u=\sigma^A,Y),
  \eeq
where $P_m^{\rmL}(0)=\delta_{m1}$, so that $Z^{\rmL}(u,Y=0)=u$.

\texttt{(b)} {\em Projectile (dipole) rest frame} (i.e., target
evolution). In this case, $Y_0=0$ and Eq.~(\ref{eq-Stot}) implies:
  \beq\label{eq-SPRF}
  \avg{S}_{Y}^{\rmR} = \sum_{n}P_n^{\rmR}(Y)\,
  \,\sigma^{n}\,=\,Z^{\rmR}(u=\sigma,Y),
  \eeq
where this time $P_n^{\rmR}(0)=\delta_{nA}$, implying
$Z^{\rmR}(u,Y=0)=u^A$.

We have related here $\avg{S}_{Y}$ to the generating functional
$Z(u,Y)$, cf. Eq.~(\ref{eq-Sm}), since as shown in Ref.
\cite{KozLev06} the latter is the quantity which is most
conveniently evaluated in the context of the reaction model.
Specifically, it is first straightforward to use
Eq.~(\ref{eq-dPdYrec}) with the vertices $f_n$ and $g_n$ from
Eq.~(\ref{eq-reacdiff}) to deduce the following evolution equation
for $Z(u,Y)$ in the reaction model (to be compared to
Eq.~(\ref{eq-dZdY})) :
 \beq\label{eq-dZdYreaction}
   \frac{\dif Z}{\dif Y} \,= \,u(1-u)\left\{ -\alpha\,
 \frac{\del Z}{\del u}\,+\,\frac{\beta}{2}
 \,\frac{\del^2 Z}{\del u^2}\right\}.
  \eeq
This equation is the starting point of the analysis in Ref.
\cite{KozLev06}. As shown there, and it can be also easily verified
by inspection, this equation admits the following, non--trivial,
fixed point
 \beq\label{eq-Zasreaction}
 Z_{\infty}(u)\,=\,\frac{\rme^{u/\tau_0} - 1}{\rme^{1/\tau_0} -
 1}\,,\qquad(\tau_0\equiv \beta/2\alpha),
 \eeq
which, as anticipated by our notations, is the same as the
high--energy limit of the {\em physical} generating functional
(i.e., the solution to Eq.~(\ref{eq-dZdYreaction}) with physical
initial conditions) : $Z(u,Y\to\infty)\to Z_{\infty}(u)$ for {\em
any initial conditions}. (This is demonstrated in Ref.
\cite{KozLev06}.) This should be contrasted to the corresponding
behaviour of the model studied by us here, where we have seen that
$Z(u,Y)$ vanishes exponentially when $Y\to\infty$ for any $u < 1$
(cf. Eqs.~(\ref{eq-Zsmallu}) and (\ref{eq-ZYlimit})).

By substituting Eq.~(\ref{eq-Zasreaction}) in Eqs.~(\ref{eq-STRF})
and (\ref{eq-SPRF}), it is easy to compute the asymptotic values of
the (average) $S$--matrix in the two Lorentz frames under
consideration. Explicitly, one obtains
 \beq\label{eq-SPRFas}
  \avg{S}_{\infty}^{\rmL} \,=\,
  \frac{\rme^{\sigma^A/\tau_0} - 1}{\rme^{1/\tau_0} -1}\,\simeq\,
  \rme^{-(1-\sigma^A)/\tau_0} \,-\,\rme^{-1/\tau_0}\,,
  \eeq
where the second, approximate, equality holds since $\tau_0\ll 1$.
The quantity $1-\sigma^A$ can be recognized as the original value of
the scattering amplitude at $Y=0$: $\avg{T}_{0}= 1-
\avg{S}_0=1-\sigma^A$. It is not hard to show that
$\avg{S}_{\infty}^{\rmL} < \avg{S}_0 = \sigma^A$ for any positive
value of $\tau_0$ and any value of $\sigma^A$ such that
$0<\sigma^A<1$. Thus, in this case, one reaches the natural
conclusion that the value of the $S$-matrix becomes smaller at high
energy.

On the other hand,
 \beq\label{eq-STRFas}
  \avg{S}_{\infty}^{\rmR} \,=\,
  \frac{\rme^{\sigma/\tau_0} - 1}{\rme^{1/\tau_0} -1}\,\simeq\,
  \rme^{-\tau/\tau_0} \,-\,\rme^{-1/\tau_0}\,,
  \eeq
where the exponent $\tau/{\tau_0}$ is of order one. Clearly, for any
$A> 1$, the two above results are different from each other, namely,
one has $1-\sigma^A > 1-\sigma\equiv\tau$ and hence
$\avg{S}_{\infty}^{\rmL} < \avg{S}_{\infty}^{\rmR}$.

Moreover, when $A$ is large enough, the results above predict that
the asymptotic $S$--matrix at high energy may become {\em larger}
than the original $S$--matrix $\avg{S}_0=\sigma^A$ at $Y=0$. Indeed,
by using $|\ln\sigma|\simeq\tau$, one can easily check that
$\avg{S}_{\infty}^{\rmR} > \sigma^A$ as soon as $A> 1/\tau_0$.

\section{A useful sum}
\setcounter{equation}{0}

In this Appendix we calculate the following sum (with
$\sigma=1-\tau$)
  \beq\label{eq-sum1app}
   \mcal{S}_1(\tau) \equiv \tau \sum_{n=1}^{\infty}
   \frac{\sigma^n}{1-\sigma^n}.
  \eeq
Up to the $\tau$ prefactor, this is equal to the double sum
$\sum_{m,n=1}^{\infty} \sigma^{mn}$, which cannot be written in
terms of known functions. Nevertheless, we can obtain an analytic
expression for $\tau \ll 1$, by keeping the first few terms in a
series expansion. Here we shall work up to order $\order{\tau}$. (Of
course, the summation can be easily evaluated numerically, since it
is rapidly convergent for any $\sigma< 1$.)

One can convert the sum into an integral by making use of the
Euler--McLaurin summation formula, which may be written as
  \beq\label{eq-eulermc}
   \sum_{n=1}^{\infty} h_n =
   \int\limits_{1}^{\infty}\dif n\, h(n)
   + \frac{1}{2}\, h(1)
   + 2 \sum_{k=1}^{\infty}
   \frac{(-1)^k \zeta(2 k)}{(2 \pi)^{2 k}}\,h^{(2 k -1)}(1).
  \eeq
With $h(n) = \tau \sigma^n/(1-\sigma^n)$, it is straightforward to
calculate the integral and obtain
  \beq\label{eq-sum1first}
   \int\limits_{1}^{\infty} \dif n\,
   \frac{\tau \sigma^n}{1-\sigma^n} =
   \frac{\tau \ln\tau}{\ln(1-\tau)} =
   \ln\frac{1}{\tau}-\frac{1}{2}\,\tau\ln\frac{1}{\tau}
   +\order{\tau^2}.
  \eeq
The integral contains the dominant contribution for small $\tau$,
which is equal to $\ln(1/\tau)$. It is trivial to obtain the
contribution of the second term in Eq.~(\ref{eq-eulermc}) which is
  \beq\label{eq-sum1second}
   \frac{1}{2}\,h(1) = \frac{1}{2} - \frac{\tau}{2}.
  \eeq
For the contribution of the third term in Eq.~(\ref{eq-eulermc}) we
need the $(2k\!-\!1)$-th derivative of $h(n)$ at $n=1$ which, to the
order of interest, reads
  \beq\label{eq-hder}
   h^{(2k-1)}(1) = -\Gamma(2k)\left(1-\frac{\tau}{2}\right)
   + \order{\tau^2}.
  \eeq
Notice that we have to sum a series whose terms are growing
factorially. However, the terms have alternating signs and the
series is Borel summable. Thus we can write the third term as
  \beq\label{eq-sum1third}
   && -2 \left(1-\frac{\tau}{2}\right)
   \sum_{k=1}^{\infty} \frac{(-1)^k\, \zeta(2 k)\,\Gamma(2 k)}
   {(2 \pi)^{2 k}}
   \nn &=&
   -2 \left(1-\frac{\tau}{2}\right)
   \int\limits_{0}^{\infty} \frac{\dif b}{b}\,\rme^{-b}
   \sum_{1}^{\infty} (-1)^k\, \zeta(2 k)
   \left(\frac{b}{2 \pi} \right)^{2 k}
   \nn &=&
   \left(1-\frac{\tau}{2}\right)
   \int\limits_{0}^{\infty} \frac{\dif b}{b}\,\rme^{-b}\,
   \left[\frac{b}{2}\, \coth \left(\frac{b}{2} \right) -1\right] =
   \left(\gamma_{\rm E}-\frac{1}{2} \right)
   \left(1-\frac{\tau}{2}\right).
  \eeq
Adding the three contributions obtained in
Eqs.~(\ref{eq-sum1first}), (\ref{eq-sum1second}) and
(\ref{eq-sum1third}) we finally arrive at
  \beq\label{eq-sum1total}
   \mcal{S}_1(\tau) =
   \ln\frac{1}{\tau}
   + \gamma_{\rm E}
   - \frac{1}{2}\,\tau\ln\frac{1}{\tau}
   -\frac{1+ 2 \gamma_{\rm E}}{4}\,\tau
   +\order{\tau^2}.
  \eeq
Similarly one can obtain the other sums encountered in
Eq.~(\ref{eq-sumr}). One finds
  \beq\label{eq-sumrapp}
   \mcal{S}_r(\tau) \equiv
   \tau^r \sum_{n=1}^\infty \frac{1- (1-\sigma^n)^r}{(1-\sigma^n)^r}
   = \zeta(r) + \order{\tau}
   \quad \text{for} \quad r \geq 2.
  \eeq


\begin{thebibliography}{10}

\bibitem{AMSalam95}
A.H.~Mueller and G.P.~Salam, {\it Nucl. Phys.} {\bf B475} (1996)
293.

\bibitem{AM94}
A.H.~Mueller, {\it Nucl. Phys.} {\bf B415} (1994) 373; A.H. Mueller,
B. Patel,
  {\it Nucl. Phys.} {\bf B425} (1994) 471.

\bibitem{AM95}
A.H.~Mueller, {\it Nucl. Phys.} {\bf B437} (1995) 107.

\bibitem{Salam95}
G.P.~Salam, {\it Nucl. Phys.} {\bf B449} (1995) 589; {\it Nucl.
Phys.} {\bf
  B461} (1996) 512.

\bibitem{B}
I.~Balitsky, {\it Nucl.\ Phys.}\ {\bf B463} (1996) 99; {\it Phys.
Lett.} {\bf
  B518} (2001) 235; {\it ``High-energy QCD and Wilson lines''},
  arXiv:hep-ph/0101042.

\bibitem{K}
Yu.V.~Kovchegov, {\it Phys. Rev.} {\bf D60} (1999) 034008; {\it
ibid.} {\bf
  D61} (1999) 074018.

\bibitem{JKLW}
J.~Jalilian-Marian, A.~Kovner, A.~Leonidov and H.~Weigert, {\it
Nucl.\ Phys.}\ {\bf B504} (1997) 415; {\it Phys.\ Rev.}\ {\bf D59}
(1999) 014014;
  J.~Jalilian-Marian, A.~Kovner and H.~Weigert,
  {\it Phys.\ Rev.}\ {\bf D59}
  (1999) 014015; A. Kovner, J. G. Milhano and H. Weigert,
  {\it Phys. Rev.} {\bf D62} (2000) 114005.

\bibitem{W}
H.~Weigert, {\it Nucl. Phys.} {\bf A703} (2002) 823.

\bibitem{CGC}
E.~Iancu, A.~Leonidov and L.~McLerran, {\it Nucl. Phys.}~{\bf A692}
(2001) 583;
  {\it Phys. Lett.} {\bf B510} (2001) 133;
  E.~Ferreiro, E.~Iancu, A.~Leonidov and L.~McLerran,
  {\it Nucl. Phys.} {\bf A703} (2002) 489.

\bibitem{PATH}
J.-P.~Blaizot, E.~Iancu and H.~Weigert, {\it Nucl.~Phys.~}{\bf A713}
(2003)
  441.

\bibitem{IM031}
E.~Iancu and A.H.~Mueller, {\it Nucl.\ Phys.}\ {\bf A730} (2004)
460.

\bibitem{IM032}
E.~Iancu and A.H.~Mueller, {\it Nucl.\ Phys.}\ {\bf A730} (2004)
494.

\bibitem{MP03}
S.~Munier and R.~Peschanski, {\it Phys. Rev. Lett.} {\bf 91} (2003)
232001;
  {\it Phys.\ Rev.}\ {\bf D69} (2004) 034008;
  {\it ibid.} {\bf D70} (2004) 077503.

\bibitem{MS04}
A.H.~Mueller and A.I.~Shoshi, {\it Nucl.\ Phys.}\ {\bf B692} (2004)
175.

\bibitem{IMM04}
E.~Iancu, A.H.~Mueller and S.~Munier, {\it Phys.~Lett.~}{\bf B606}
(2005) 342.

\bibitem{LL04}
E.~Levin and M.~Lublinsky, {\it Phys. Lett.} {\bf B607} (2005) 131.

\bibitem{IT04}
E.~Iancu and D.N.~Triantafyllopoulos, {\it Nucl.~Phys.~}{\bf A756}
(2005) 419.

\bibitem{IT05}
E.~Iancu and D.N.~Triantafyllopoulos, {\it Phys.~Lett.~}{\bf B610}
(2005) 253.

\bibitem{MSW05}
A.H.~Mueller, A.I.~Shoshi, and S.M.H.~Wong, {\it Nucl.~Phys.~}{\bf
B715} (2005)
  440.

\bibitem{LL05}
E.~Levin and M.~Lublinsky, {\it Nucl.~Phys.~}{\bf A763} (2005) 172.

\bibitem{KL05}
A.~Kovner and M.~Lublinsky, {\it Phys.~Rev.~}{\bf D71} (2005)
085004.

\bibitem{KL3}
A.~Kovner and M.~Lublinsky, {\it Phys.~Rev.~Lett.~}{\bf 94} (2005)
181603.

\bibitem{BIIT05}
J.-P.~Blaizot, E.~Iancu, K.~Itakura, and D.N.~Triantafyllopoulos,
{\it
  Phys.~Lett.~}{\bf B615} (2005) 221.

\bibitem{BREM}
Y. Hatta, E. Iancu, L. McLerran, A. Stasto, D.N.~Triantafyllopoulos,
{\it  Nucl. Phys.} {\bf A764} (2006) 423.

\bibitem{Balit05}
I. Balitsky, {\it Phys. Rev.} {\bf D72} (2005) 074027.

\bibitem{IST05}
E.~Iancu, G. Soyez and D.N.~Triantafyllopoulos, {\it Nucl. Phys.}
{\bf A768}
  (2006) 194.

\bibitem{Gardiner}
C.W. Gardiner, {\it Handbook of Stochastic Methods},  Springer,
Berlin, 2004.

\bibitem{Saar}
For a recent review, see W.~Van Saarloos, {\it Phys. Rep.} {\bf 386}
(2003) 29.

\bibitem{Braun05}
M.~Braun, {\it Phys.~Lett.~}{\bf B483} (2000) 115; {\it ``Conformal
invariant
  equations for nucleus-nucleus scattering in perturbative QCD with
  $N_c\to\infty$''}, arXiv:hep-ph/0504002.

\bibitem{GS05}
G. Soyez, {\it Phys. Rev.} {\bf D72} (2005) 016007.

\bibitem{EGBM05}
R. Enberg, K. Golec--Biernat, and S. Munier, {\it Phys. Rev.} {\bf
D72} (2005)
  074021.

\bibitem{Stasto05}
P. Rembiesa and A.M. Stasto, {\it Nucl. Phys.} {\bf B725} (2005)
251.

\bibitem{KL5}
A.~Kovner and M.~Lublinsky, {\it Nucl. Phys.} {\bf A767} (2006) 171.

\bibitem{SX05}
A. I. Shoshi and B.-W. Xiao, {\it Phys. Rev.} {\bf D73} (2006)
094014; {\it ``Diffractive dissociation including pomeron loops in
zero transverse dimensions''}, arXiv:hep-ph/0605282.

\bibitem{KozLev06}
M. Kozlov and E. Levin, {\it ``Solution for the BFKL Pomeron
Calculus in zero
  transverse dimensions''} hep-ph/0604039.

\bibitem{MV}
L.~McLerran and R.~Venugopalan, {\it Phys.\ Rev.}\ {\bf D49} (1994)
2233; {\it
  ibid.} {\bf 49} (1994) 3352; {\it ibid.} {\bf 50} (1994) 2225.

\bibitem{EdiCGC}
E.~Iancu, A.~Leonidov and L.~McLerran, {\it ``The Colour Glass
Condensate: An
  Introduction''}, arXiv:hep-ph/0202270. Published in
  {\it QCD Perspectives on Hot and Dense Matter},
  Eds. J.-P.~Blaizot and E.~Iancu, NATO Science Series,
  Kluwer, 2002;\\ E.~Iancu and R.~Venugopalan, {\it ``The Color Glass
  Condensate and High Energy Scattering in QCD''},
  arXiv:hep-ph/0303204.
  Published in {\it Quark-Gluon Plasma 3},
  Eds. R.C.~Hwa and X.-N.~Wang, World Scientific, 2003.

\bibitem{SCALING}
E.~Iancu, K.~Itakura, and L.~McLerran, {\it Nucl. Phys.} {\bf A708}
(2002) 327.

\bibitem{MT02}
A.H.~Mueller and D.N.~Triantafyllopoulos, {\it Nucl. Phys.} {\bf
B640} (2002)
  331.

\bibitem{DT02}
D.N.~Triantafyllopoulos, {\it Nucl. Phys.} {\bf B648} (2003) 293.

\bibitem{HIMST06}
Y.~Hatta, E.~Iancu, C. Marquet, G. Soyez, and
D.N.~Triantafyllopoulos, {\it
  Nucl. Phys.} {\bf A773} (2006) 95.

\bibitem{GLUON}
E.~Iancu, C. Marquet, and G. Soyez, {\it ``Forward gluon production
in
  hadron--hadron scattering with Pomeron loops''}, arXiv:hep-ph/0605174.

\bibitem{AM99}
A. H. Mueller, {\it Nucl. Phys.} {\bf B558} (1999) 285.

\bibitem{SAT}
E.~Iancu and L.~McLerran, {\it Phys. Lett.} {\bf B510} (2001) 145.

\bibitem{AM02}
A. H. Mueller, {\it Nucl. Phys.} {\bf B643} (2002) 501.

\bibitem{GAUSS}
E.~Iancu, K.~Itakura, and L.~McLerran, {\it Nucl. Phys.} {\bf A724}
(2003) 181.

\bibitem{BFKL}
L.N.~Lipatov, {\it Sov.\ J.\ Nucl.\ Phys.}\,{\bf 23} (1976) 338;
 E.A.~Kuraev, L.N.~Lipatov and V.S.~Fadin,
 {\it Zh. Eksp. Teor. Fiz} {\bf 72}, 3 (1977)
  ({\it Sov. Phys. JETP }{\bf 45} (1977) 199); Ya.Ya.~Balitsky and
  L.N.~Lipatov, {\it Sov.\ J.\ Nucl.\ Phys.} {\bf 28} (1978) 822.

\bibitem{B1}
I.~Balitsky, {\it Phys. Rev. Lett.} {\bf 81} (1998) 2024; {\it
Phys.\ Rev.}\
  {\bf D60} (1999) 014020.

\bibitem{MMSW05}
C. Marquet, A.H. Mueller, A.I. Shoshi, and S.M.H. Wong, {\it Nucl.
Phys.} {\bf
  A762} (2005) 252.

\bibitem{HIMS05}
Y. Hatta, E. Iancu, L. McLerran, and A. Stasto, {\it Nucl. Phys.}
{\bf A762}
  (2005) 272.

\end{thebibliography}

\end{document}